\documentclass[12pt, preprint2, tighten, longabstract]{aastex}

\usepackage{rotating}
\usepackage{amssymb}
\usepackage{longtable}
\usepackage{booktabs}
\usepackage{calc}
\usepackage{hyperref}

\shorttitle{Reconnection Inflows}
\shortauthors{Savage, Holman, Reeves, Seaton, McKenzie, \& Su}

\begin{document}

\title{LOW-ALTITUDE RECONNECTION INFLOW-OUTFLOW OBSERVATIONS DURING A 2010 NOVEMBER 3 SOLAR ERUPTION}

\author{$^{1}$Sabrina L. Savage, $^{1}$Gordon Holman, $^{2}$Katharine K. Reeves, $^{3}$Daniel B. Seaton, $^{4}$David E. McKenzie, $^{1,5}$Yang Su}
\affil{$^{1}$NASA/Goddard Space Flight Center (Oak Ridge Associated Universities), 8800 Greenbelt Rd Code 671, Greenbelt, MD  20771, USA} 
\affil{$^{2}$Harvard-Smithsonian Center for Astrophysics, 60 Garden Street MS 58, Cambridge, MA  02138}
\affil{$^{3}$Royal Observatory of Belgium-SIDC, Avenue Circulaire 3, 1180 Brussels, Belgium}
\affil{$^{4}$Department of Physics, Montana State University, PO Box 173840, Bozeman, MT 59717-3840, USA}
\affil{$^{1, 5}$ Institute of Physics, University of Graz, Universitaetsplatz 5, Graz 8010, Austria}

\begin{abstract}

For a solar flare occurring on 2010 November 3, we present observations using several \textit{SDO}/AIA extreme-ultraviolet (EUV) passbands of an erupting flux rope followed by inflows sweeping into a current sheet region.  The inflows are soon followed by outflows appearing to originate from near the termination point of the inflowing motion -- an observation in line with standard magnetic reconnection models.  We measure average inflow plane-of-sky speeds to range from $\sim$150~$-$~690~km~s$^{-1}$ with the initial, high-temperature inflows being the fastest.  Using the inflow speeds and a range of Alfv\'{e}n speeds, we estimate the Alfv\'{e}nic Mach number which appears to decrease with time.  We also provide inflow and outflow times with respect to RHESSI count rates and find that the fast, high-temperature inflows occur simultaneously with a peak in the RHESSI thermal light curve.  Five candidate inflow-outflow pairs are identified with no more than a minute delay between detections.  The inflow speeds of these pairs are measured to be $\sim$10$^{2}$~km~s$^{-1}$ with outflow speeds ranging from $\sim$10$^{2}$--10$^{3}$~km~s$^{-1}$ -- indicating acceleration during the reconnection process.  The fastest of these outflows are in the form of apparently traveling density enhancements along the legs of the loops rather than the loop apexes themselves.  These flows could possibly either be accelerated plasma, shocks, or waves prompted by reconnection.  The measurements presented here show an order of magnitude difference between the retraction speeds of the loops and the speed of the density enhancements within the loops -- presumably exiting the reconnection site.  

\end{abstract}

\keywords{Magnetic reconnection --- Sun: corona --- Sun: flares --- Sun: coronal mass ejections (CMEs) --- Sun: magnetic topology --- Sun: EUV}

\section{\label{inflows101103:intro}INTRODUCTION}

Magnetic reconnection is widely accepted as an important mechanism for energy release during solar flares; however, observations of it have been indirect and/or incomplete.  Magnetic reconnection serves to reconfigure a non-potential field into a lower energy state (i.e. the basic CSHKP model:  \citeauthor{carmichael_1964} \citeyear{carmichael_1964}; \citeauthor{sturrock_1968} \citeyear{sturrock_1968}; \citeauthor{hirayama_1974} \citeyear{hirayama_1974}; \citeauthor{kopp-pneuman_1976} \citeyear{kopp-pneuman_1976}). This reconfiguration is expected to occur at the onset of a flare and continue throughout the decay phase, but due to observational constraints such as dynamic range, saturation from footpoints and post-flare loops, and cadence, observing reconnection directly is most feasible during the decay phase when the reconnection sites are assumed to be well above the overwhelmingly bright footpoints and post-flare loops.  

\begin{figure*}[!ht] 
\begin{center}

\framebox{
\includegraphics[width=.32\textwidth]{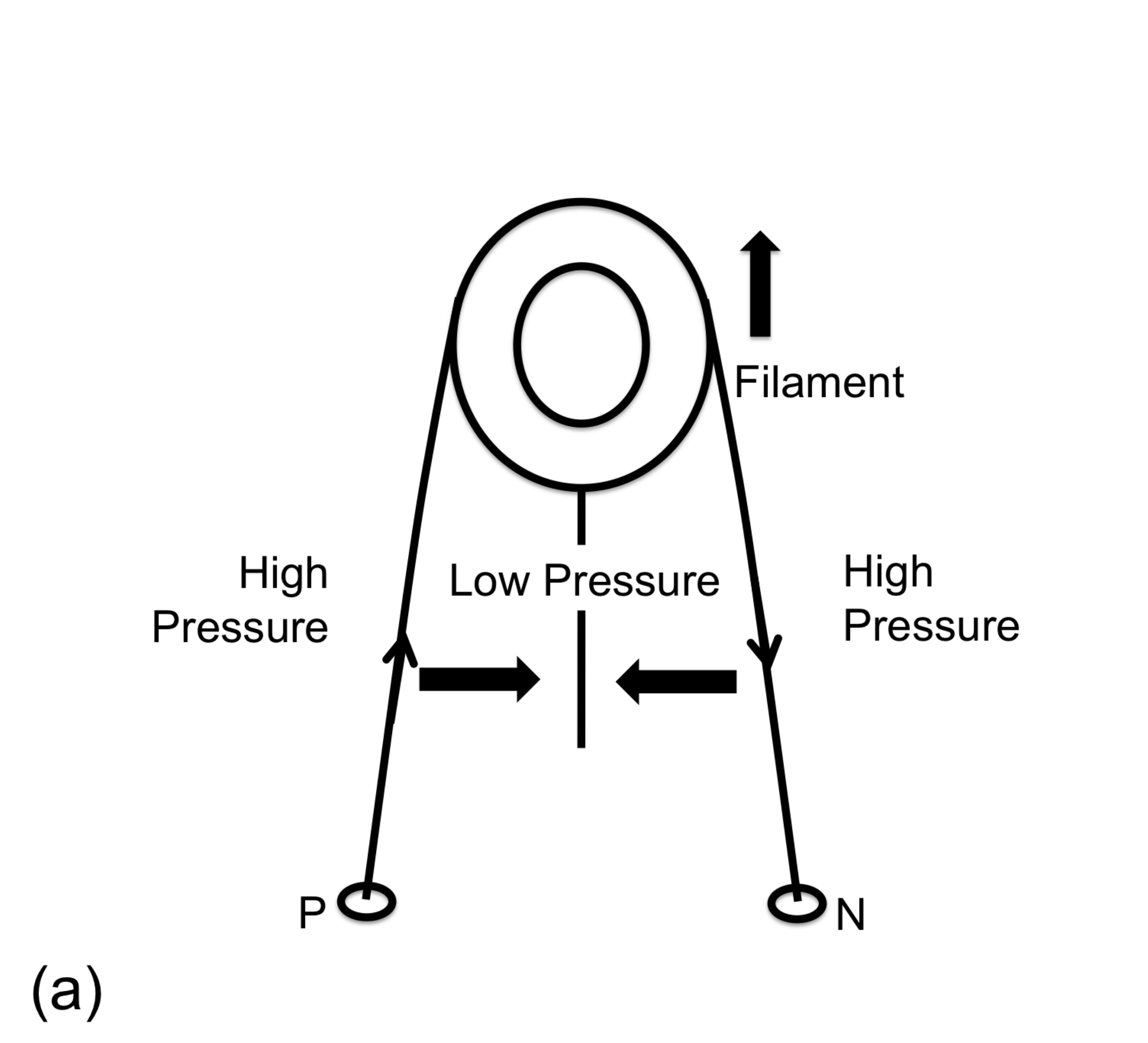}
\includegraphics[width=.32\textwidth]{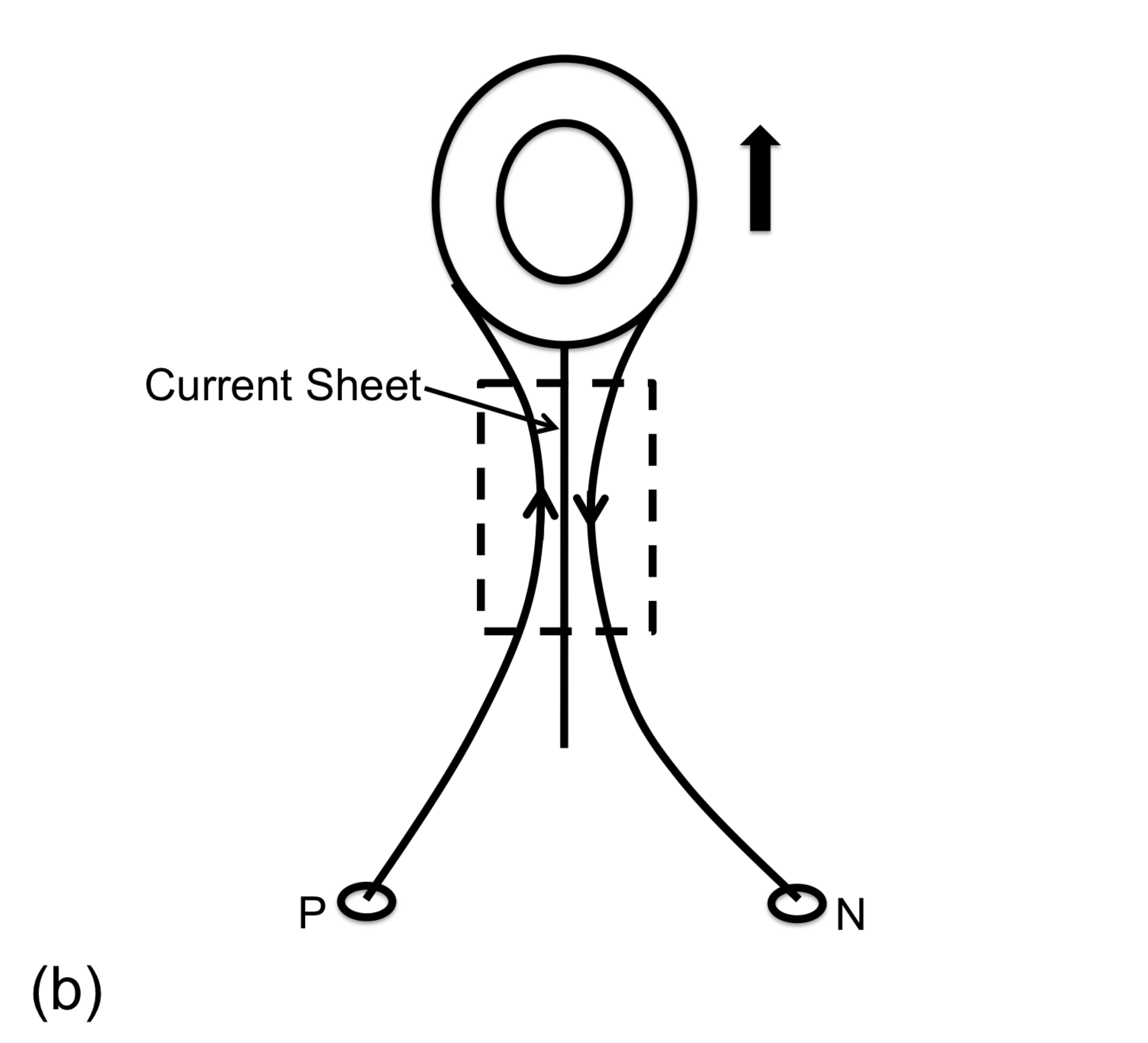}
\includegraphics[width=.32\textwidth]{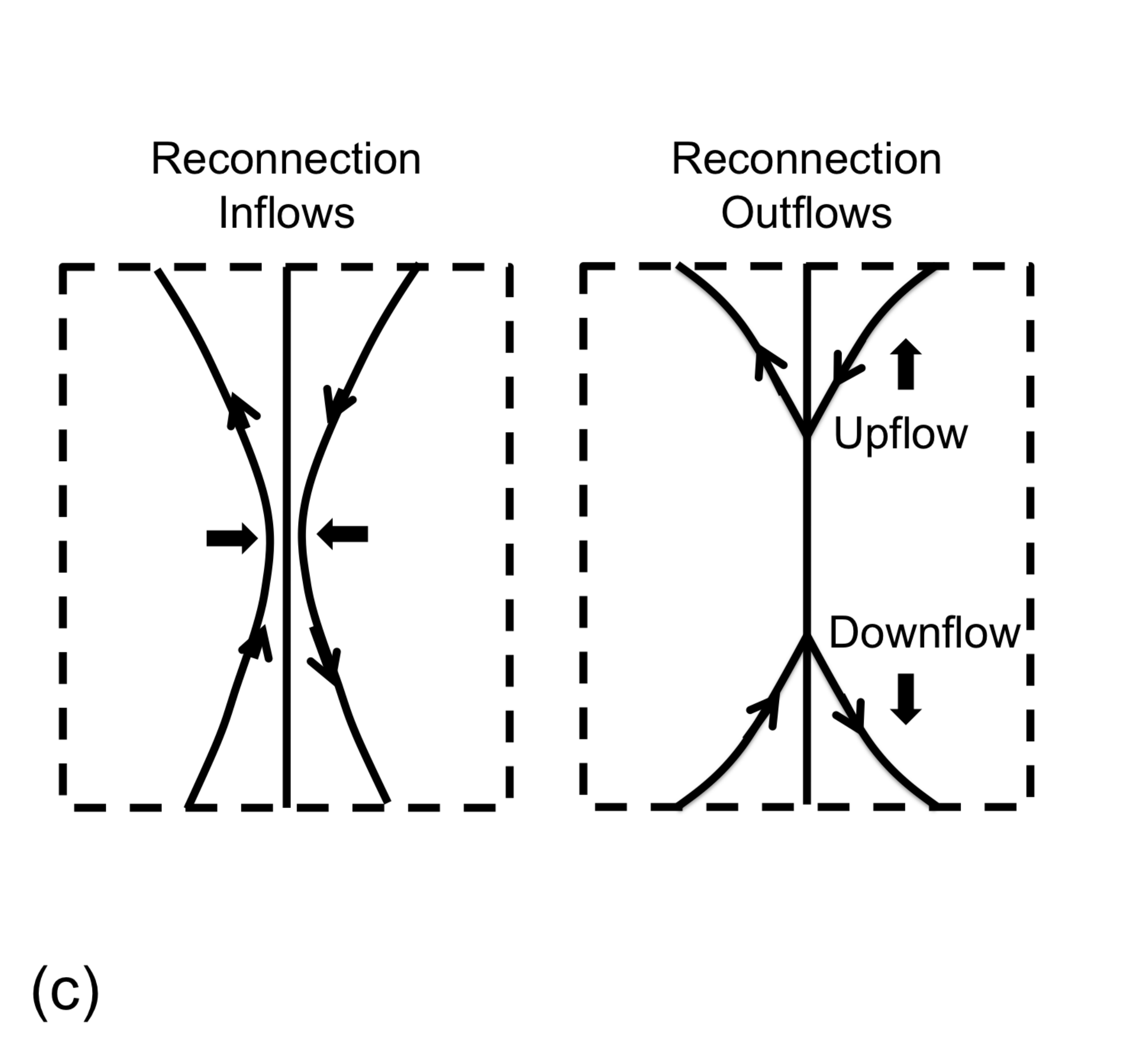}
}

\caption{Cartoon depiction of a basic 2-D reconnection scenario.}
\label{reconnection}
\end{center}
\end{figure*}

A basic scenario for post-flare reconnection can be described as follows (e.g. \citeauthor{lin-forbes_2000} \citeyear{lin-forbes_2000}; \citeauthor{forbes-acton_1996} \citeyear{forbes-acton_1996}):  1) A flux rope erupts due to a loss of equilibrium and travels out into the corona as a coronal mass ejection (CME).\footnote{As noted in \cite{sui-holman-dennis_2004}, there are several models describing the eruption itself including tether-cutting \citep{moore-roumeliotis_1992}, kink instability \citep{cheng_1977}, flux rope instability \citep{forbes-priest_1995}, magnetic breakout (\citeauthor{macneiceEA_2004} \citeyear{macneiceEA_2004}; \citeauthor{antiochos-devore-klimchuk_1999} \citeyear{antiochos-devore-klimchuk_1999}), and quadruple magnetic source \citep{hiroseEA_2001}.  We will focus on the events occurring after the flux rope is initially released.}  2)  Because of magnetic pressure imbalances in the low beta corona, field lines of opposite polarity are swept together in the wake of the erupted flux rope.  3)  A current sheet forms between these opposing polarities.  4)  Via localized regions of high resistivity in the current sheet or tearing of the current layer, field lines alongside the polarity inversion reconnect to form magnetic loops perpendicular to the current sheet (e.g. \citeauthor{bartaEA_2011a} \citeyear{bartaEA_2011a}; \citeauthor{linton-longcope_2006} \citeyear{linton-longcope_2006}).  The reconfigured fields are propelled in opposite directions via magnetic tension.  The downward-directed field forms the post-eruption arcade while the upward-directed field travels outward along with the flux rope.  (See Figure~\ref{reconnection} for a 2-D cartoon depiction of this scenario.)  Guide fields parallel to the current sheet introduce 3-D effects such as patchy reconnection and sheared outflows (\citeauthor{guidoni-longcope_2011} \citeyear{guidoni-longcope_2011}; \citeauthor{longcope-guidoni-linton_2009} \citeyear{longcope-guidoni-linton_2009}).

The field lines swept into the current sheet region are referred to as reconnection \textit{inflows} while the reconfigured loops are termed reconnection \textit{outflows}.  Outflows are naturally sub-divided into downflows and upflows, depending on whether they are moving towards (down) or away (up) from the solar surface (Figure~\ref{reconnection} (c)).  Inflow observations greater than 1.5~R$_{\odot}$ above the solar surface have previously been reported using \textit{SOHO}/LASCO (e.g. \citeauthor{sheeley-warren-wang_2007} \citeyear{sheeley-warren-wang_2007}).  Inflows nearer the surface ($\sim$~10$^{5}$~km) have been reported with \textit{SOHO}/EIT (with supporting observations from \textit{Yohkoh}/SXT and \textit{SOHO}/MDI) (e.g. \citeauthor{narukage-shibata_2006} \citeyear{narukage-shibata_2006}; \citeauthor{yokoyamaEA_2001} \citeyear{yokoyamaEA_2001}) and \textit{SOHO}/UVCS (\citeauthor{linEA_2005} \citeyear{linEA_2005}).  Various instruments spanning a broad range of wavelengths have been used to detect and measure possible reconnection outflows in the form of supra-arcade downflows (SADs) (e.g., \citeauthor{savage-mckenzie-reeves_2012} \citeyear{savage-mckenzie-reeves_2012}; \citeauthor{savage-mckenzie_2011} \citeyear{savage-mckenzie_2011}; \citeauthor{mckenzie-savage_2011} \citeyear{mckenzie-savage_2011}; \\ \citeauthor{sheeley-wang_2007} \citeyear{sheeley-wang_2007}; \citeauthor{asaiEA_2004} \citeyear{asaiEA_2004}; \citeauthor{sheeley-warren-wang_2004} \citeyear{sheeley-warren-wang_2004}; \citeauthor{mckenzie_2000} \citeyear{mckenzie_2000}; \citeauthor{mckenzie-hudson_1999} \citeyear{mckenzie-hudson_1999}), supra-arcade downflowing loops (SADLs) (e.g., \citeauthor{savage-mckenzie_2011} \citeyear{savage-mckenzie_2011}; \citeauthor{savageEA_2010} \citeyear{savageEA_2010}), jets (e.g. \citeauthor{wang-sui-qiu_2007} \citeyear{wang-sui-qiu_2007}), and plasmoids (e.g. \citeauthor{takasaoEA_2012} (\citeyear{takasaoEA_2012}; \citeauthor{milliganEA_2010} \citeyear{milliganEA_2010}; \citeauthor{lin-cranmer-farrugia_2008} \citeyear{lin-cranmer-farrugia_2008}; \citeauthor{ohyama-shibata_2008} \citeyear{ohyama-shibata_2008}; \\ \citeauthor{rileyEA_2007} \citeyear{rileyEA_2007}; \citeauthor{linEA_2005} \citeyear{linEA_2005}; \citeauthor{sui-holman-dennis_2004} \citeyear{sui-holman-dennis_2004}; \citeauthor{koEA_2003} \citeyear{koEA_2003}).  

\begin{figure*}[!ht] 
\begin{center}

\includegraphics[height=450pt]{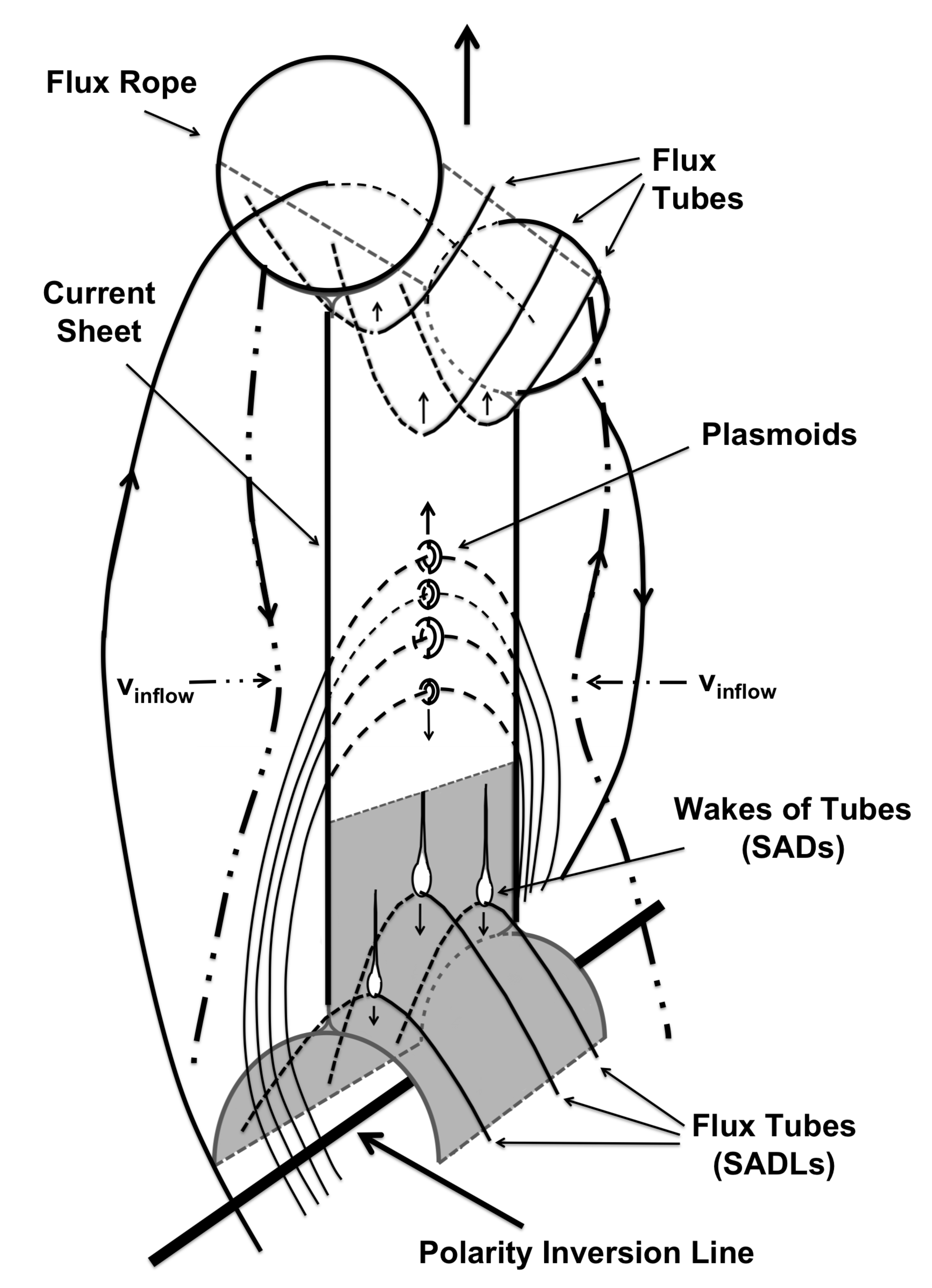}

\caption{Schematic depiction of a basic reconnection scenario, post initial flux rope formation and release, indicating the general organization of the magnetic field of the various components.  Field lines of opposite polarities (heavy dashed, double-dot lines), stretched in the wake of the erupting flux rope, flow inwards due to magnetic pressure imbalance.  The field lines reconnect across the current sheet to form outflowing flux tubes (Figure~\ref{reconnection}) while plasmoids form in the current sheet through instabilities.  Wakes (SADs) are formed as the flux tubes retract through hot plasma in the current sheet region above the arcade -- particularly in the late phase of the event (\textit{otherwise, only SADLs are observed}).  The solid/dashed curved lines extending through the flux rope and plasmoids represent the general plane and footpoint organization of the local magnetic field.}
\label{cartoon_all}
\end{center}
\end{figure*}

Figure~\ref{cartoon_all} provides a combined schematic depiction of flux tubes and plasmoids forming in the wake of an erupting flux rope.  \cite{ohyama-shibata_2008} show schematically how X-ray plasmoids can form in a CME current sheet (Figure 10 therein).  Similarly, \cite{daughton-roytershteyn_2010} explain that ``both primary and secondary magnetic islands correspond to extended flux ropes in 3D"; however, their associated magnetic fields are likely considerably more complex than as depicted in Figure~\ref{cartoon_all}.  Plasmoid development and evolution is beyond the scope of the current paper, but for a localized description of plasmoids -- or magnetic islands -- resulting from current sheet instabilities, we refer the reader to simulation papers such as \citeauthor{fermoEA_2011} \citeyear{fermoEA_2011}, \citeauthor{daughton-roytershteyn_2010} \citeyear{daughton-roytershteyn_2010}, and \citeauthor{drakeEA_2006} \citeyear{drakeEA_2006}.  

%The plasmoid field lines are expected to be significantly more complicated than depicted in Figure~\ref{cartoon_all} \citep{daughton-roytershteyn_2010}. 

The simplified cartoon in Figure~\ref{cartoon_all} serves to indicate the general geometry and organization of the magnetic field and related features involved with the reconnection and eruption.Ê Specifically, SADLs are loops that appear to retract quickly through the current sheet region from high in the corona until their motion ceases as they approach the potential post-eruption arcade.  These loops are considered to be post-reconnection outflows.  Meanwhile, SADs are oblong voids that have similar trajectories through the same current sheet region.  Based on prior analyses and available data, SADs were interpreted as the cross-sections of the same post-reconnection loops as seen from an orientation perpendicular to the current sheet.  (For diagrams and examples, see \citeauthor{savage-mckenzie_2011} \citeyear{savage-mckenzie_2011} -- Figure~2 therein, \citeauthor{savageEA_2010} \citeyear{savageEA_2010}, \citeauthor{sheeley-wang_2007} \citeyear{sheeley-wang_2007}, \citeauthor{tripathiEA_2007} \citeyear{tripathiEA_2007}, \citeauthor{asaiEA_2004} \citeyear{asaiEA_2004}, \citeauthor{sheeley-warren-wang_2004} \citeyear{sheeley-warren-wang_2004}, \citeauthor{mckenzie_2000} \citeyear{mckenzie_2000}, and \citeauthor{mckenzie-hudson_1999} \citeyear{mckenzie-hudson_1999}.)  Recent detailed analysis of an event occurring on 2011 October 22, however, reveals that SADs are more likely to be density depletions, or \textit{wakes}, that trail behind thin loops shrinking through emitting plasma associated with the current sheet rather than the \textit{cross-sections} of larger, evacuated flux tubes \citep{savage-mckenzie-reeves_2012}.

Most previous observations of possible reconnection sites have consisted of either inflows or outflows independently -- especially in the form of plasmoids.  To our knowledge, however, this is possibly the clearest observation of inflows directly followed by outflows along a current sheet in the form of SADLs.  Another similar limb flare observation focusing on inflows and outflowing plasmoids along a current sheet using AIA data has been reported by \citeauthor{takasaoEA_2012} (\citeyear{takasaoEA_2012}).  In this paper we present evidence for a complete reconnection observation following the 2010 November 3 C4.9 solar flare.  The line-of-sight of this eruption -- nearly parallel to the current sheet region and positioned a few degrees beyond the east limb -- allows for observations of inflows followed by outflows in the form of SADLs.  

\clearpage

\clearpage

\newpage

\section{\label{inflow101103:observations}OBSERVATIONS}

\begin{figure*}[!ht] 
\begin{center}

\includegraphics[width=.95\textwidth]{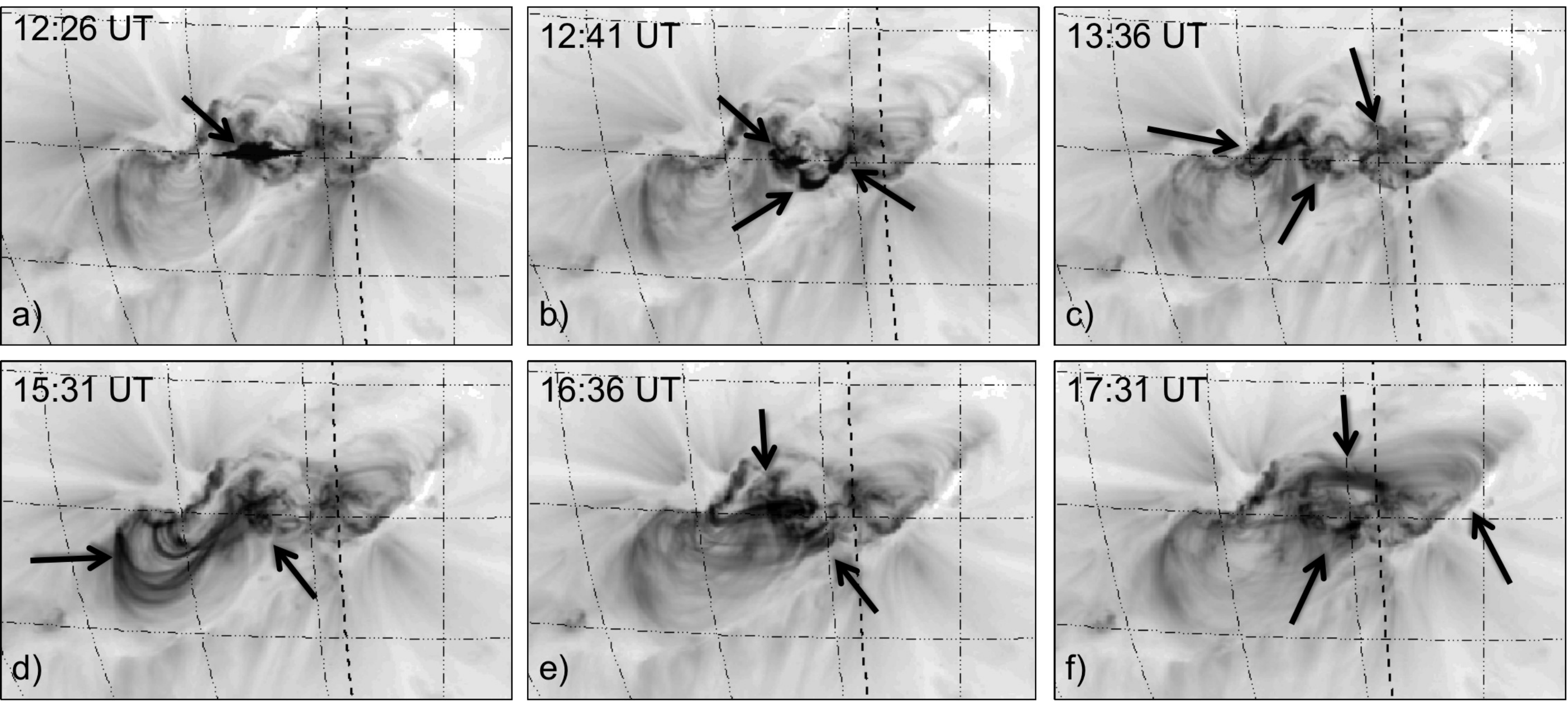}

\caption{\textit{STEREO B}/SECCHI image sequence of the event showing the complicated morphology and progression of the erupting region.  The arrows point to areas of activity.  The thick dashed longitudinal lines indicate the solar limb as viewed from Earth.  (Refer to the supplemental online material for a feature-matching movie highlighting the activity of the region throughout the event.  The Appendix contains a description of this movie.)}
\label{stereo_seq}
\end{center}
\end{figure*}

On 2010 November 3 at $\sim$12:10 UT, \textit{SDO}/AIA and \textit{STEREO-B}/SECCHI observed the eruption of a flux rope from AR 11121, which was $\sim$0--10$^{\circ}$ behind the east limb (Figure~\ref{stereo_seq}). A second, smaller eruption occurred in the same region at $\sim$12:50~UT. For a complete description of these events refer to \citet{reeves-golub_2011} and \citet{chengEA_2011}. Within the first few minutes of the eruption, a bright linear feature, which we refer to as the candidate current sheet (see Figure~\ref{orientation_images} b) and c)), appears in the AIA 211, 335, 94, and 131~$\mbox{\AA}$ channels\footnote{The plasma temperatures to which the narrow band AIA 171, 193, 211, 335, 94, and 131~$\mbox{\AA}$ bandpasses admit significant response under flaring conditions are approximately .7~MK, 6 \& 20~MK, 2-5~MK, 3~MK, 7-10~MK, and 11-14~MK, respectively (\citeauthor{odwyerEA_2010} \citeyear{odwyerEA_2010}).  Depending on the specific plasma conditions, the 193~$\mbox{\AA}$ passband may also contain significant emission at $\sim$.6~MK.} and persists until the second eruption.

Within minutes of the initial flux rope eruption, a series of flows begin retreating sunward from an apparent reconnection region along this candidate current sheet. These downflows, which take the form of both SADs and SADLs depending on the density of hot plasma above the post-flare arcade, continue for at least four hours after the initial eruption. We also observe several upflows and a number of possible disconnection events (i.e. upflows and downflows initiating simultaneously from nearly the same origin).  For the sake of brevity, we will focus on the time period between 12:10 and 12:38~UT.  During this time period, most of the outflows are observed as SADLs at 131~$\mbox{\AA}$ with a few later ones moving in from beyond the field of view appearing as SADs or plasmoids.  In the 211 and 193~$\mbox{\AA}$ bandpasses, the outflows appear more as plasma rapidly moving down loop legs.  (See Section~\ref{inflows101103:outflows} for more discussion.)  Inflows and outflows (SADs, SADLs, and upflows) continue beyond 12:38~UT.

Several hours after the onset of the event, the erupting flux rope is observed by the white-light LASCO coronagraph as a CME. Similarly, LASCO C2 images also show evidence of SADs in the apparent current sheet region later in the event.  These flows first appear approximately 18 hours after the initial eruption but within only a few hours of the passage of the CME through the LASCO field of view.  

This paper focuses on observations of inflows that are promptly followed by perpendicular outflows, indicating the possible location of a reconnection region within the candidate current sheet.  These observations occur soon after the flux rope passes through the AIA field of view. 

\subsection{\label{inflow101103:inflows}Reconnection Inflows}

We find that the inflows are most easily tracked in the 171, 193, 211, and 131~$\mbox{\AA}$ AIA bandpasses.   (A description of the manual tracking procedure used is given in Section~\ref{inflow101103:inflow_measurements}.) Note that these inflows are also observed with the 335 and 94~$\mbox{\AA}$ bandpasses; however, the flows in these channels overlap nearly completely with the chosen filter set and are not as distinctive.  In 171, 193, and 211, the inflows appear as extended bright linear features that are swept into the candidate current sheet.  In the high temperature 131 channel, the bright region directly encountering the candidate current sheet appears more localized.  (Refer to flows 2, 3, and 4 in the 131~$\mbox{\AA}$ panel of the supplemental movie accompanying this paper in the online journal.  A description of this movie can be located in the Appendix.)

While it appears that the full extent of the inflows at 131~$\mbox{\AA}$ are observable, the complete context of the inflows from the other filters is limited by the field of view (i.e.,  the 171~$\mbox{\AA}$ inflows appear to continue beyond the AIA field of view).  Because of the inability to observe the entire linear features in 171, their interpretation as stretched field lines extending to the flux rope may be questioned.  However, the difference in heights between the 131 \& 171~$\mbox{\AA}$ features is likely due to the progression of the inflows towards higher altitudes with time as the current sheet extends to greater heights, which is predicted by the standard model.  The 131~$\mbox{\AA}$ inflows occur earlier during the high energy release phase and thus lower in the field of view.  The appearance, motion, and disappearance of the inflows followed by downflows originating in the same vicinity is in line with the standard magnetic reconnection model and lends support to this interpretation.

\begin{figure*}[!ht] 
\begin{center}

\framebox{
\includegraphics[width=.32\textwidth]{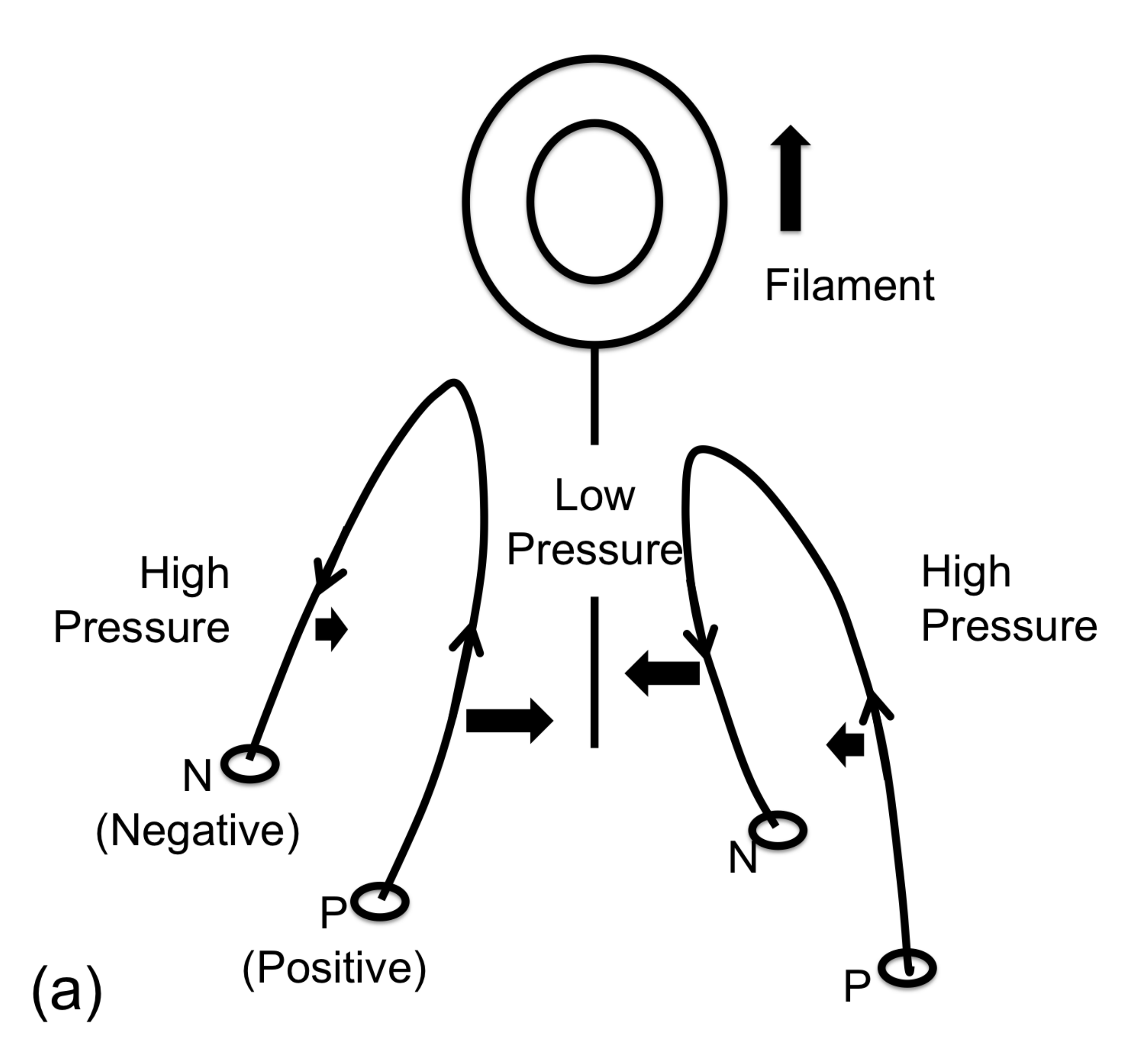}
\includegraphics[width=.32\textwidth]{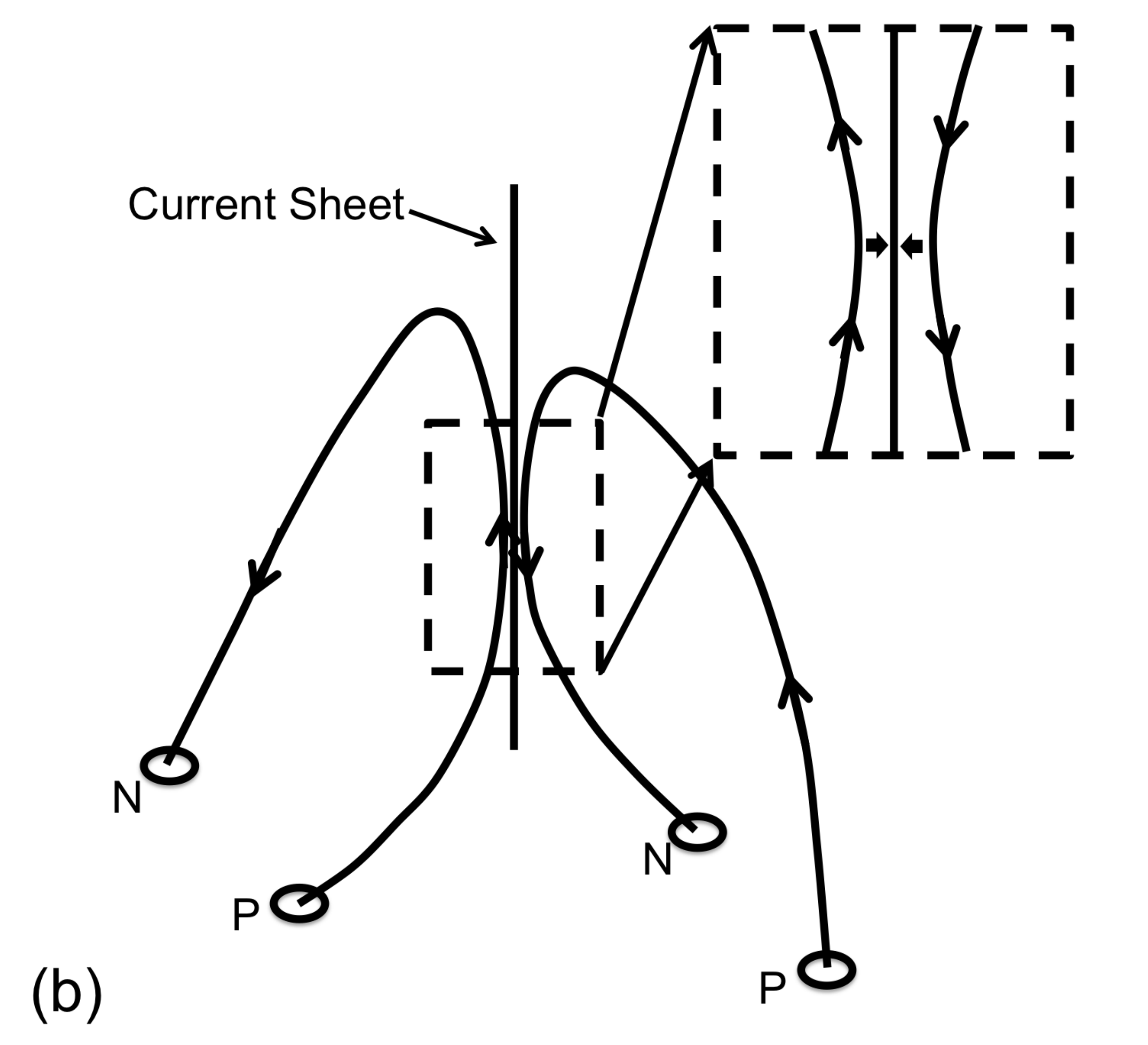}
\includegraphics[width=.32\textwidth]{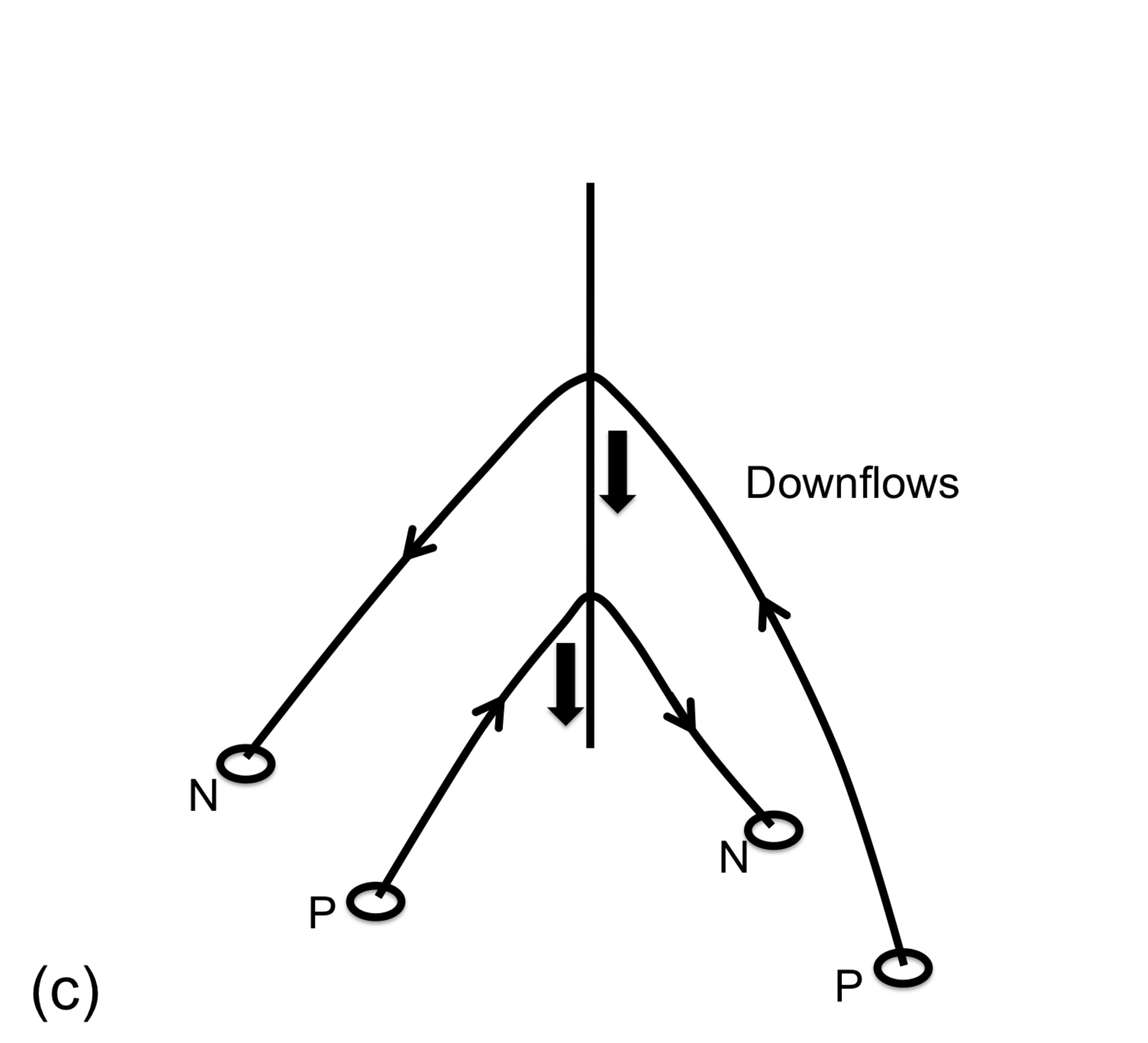}
}

\framebox{
\includegraphics[width=.32\textwidth]{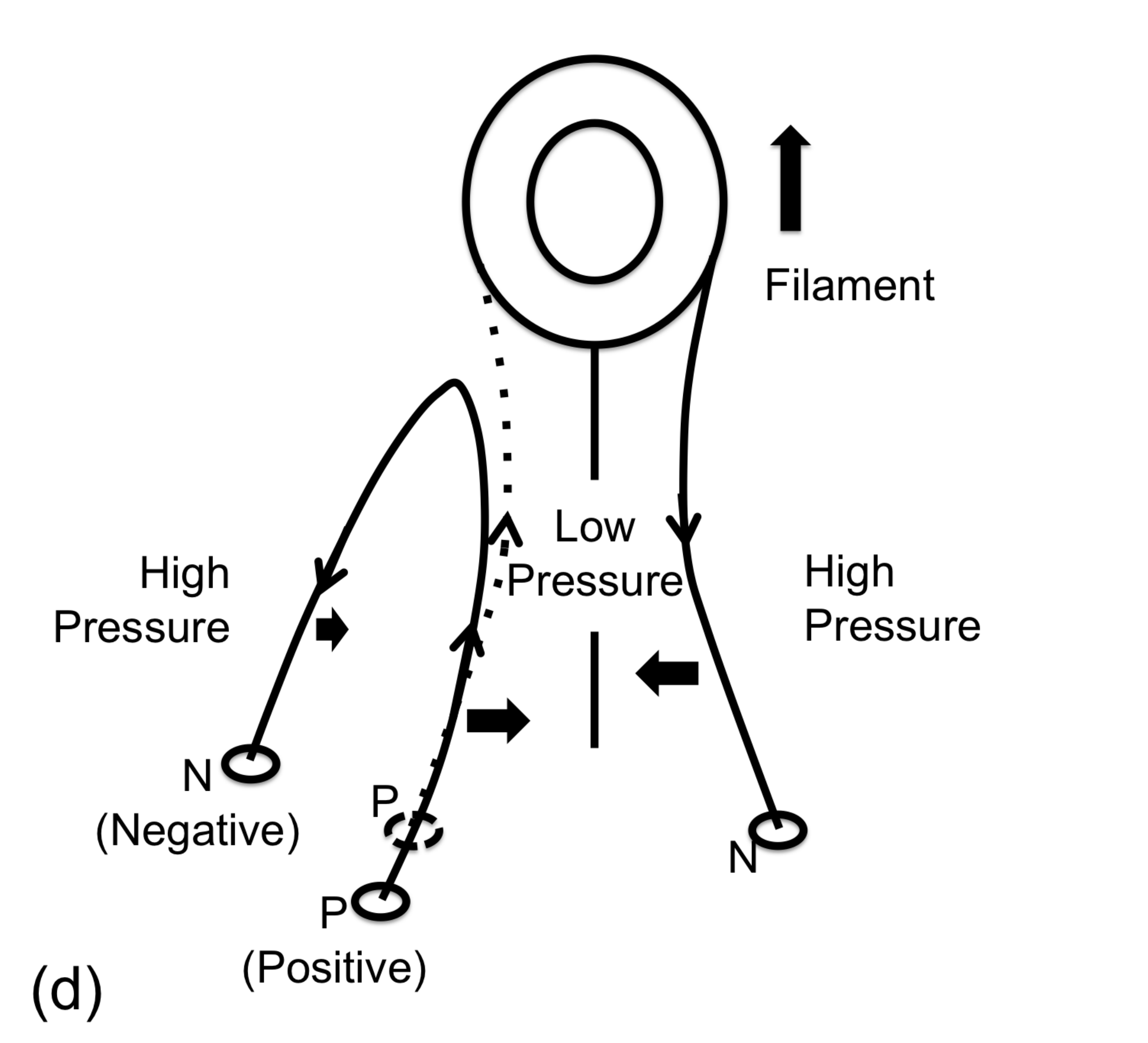}
\includegraphics[width=.32\textwidth]{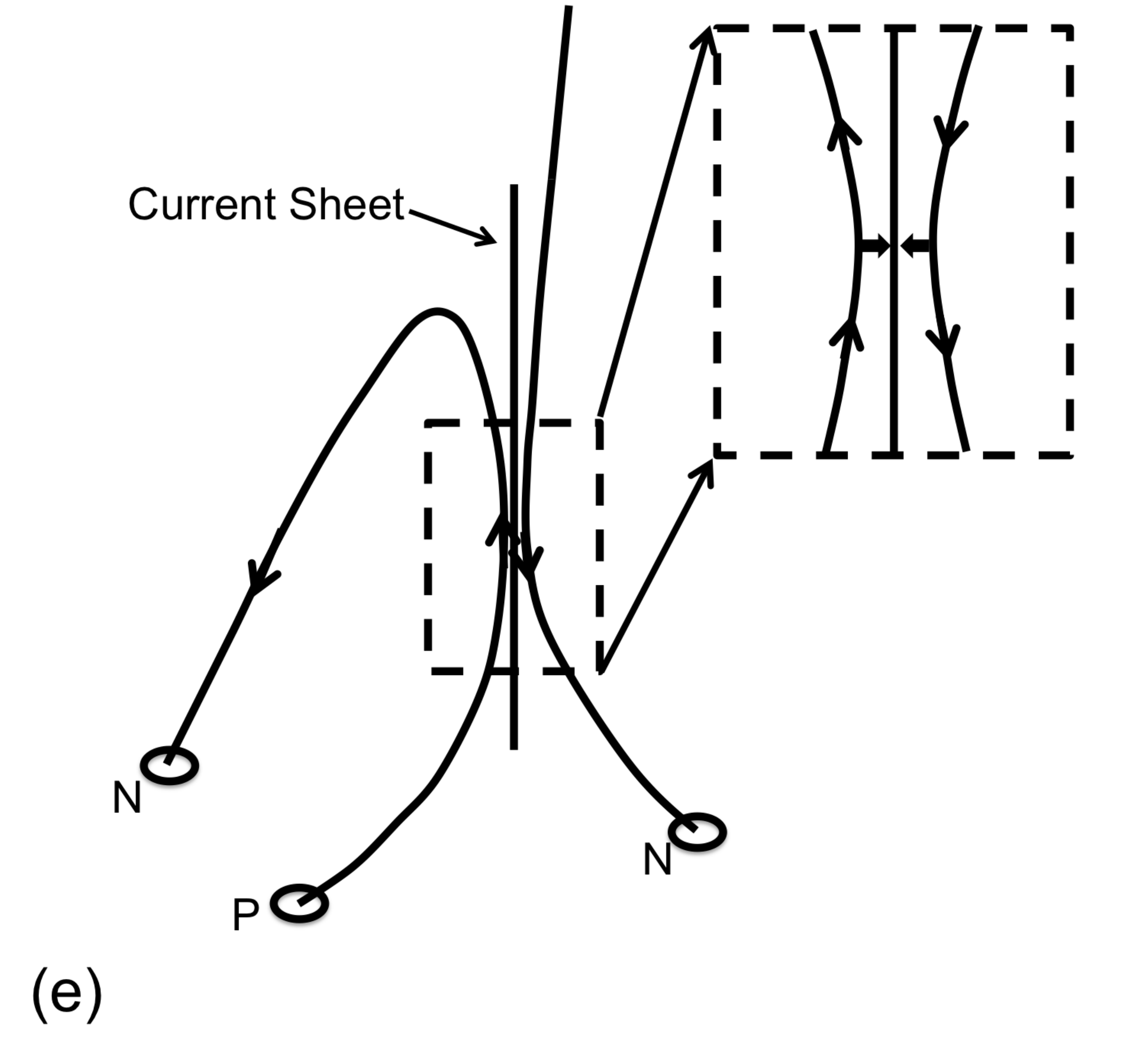}
\includegraphics[width=.32\textwidth]{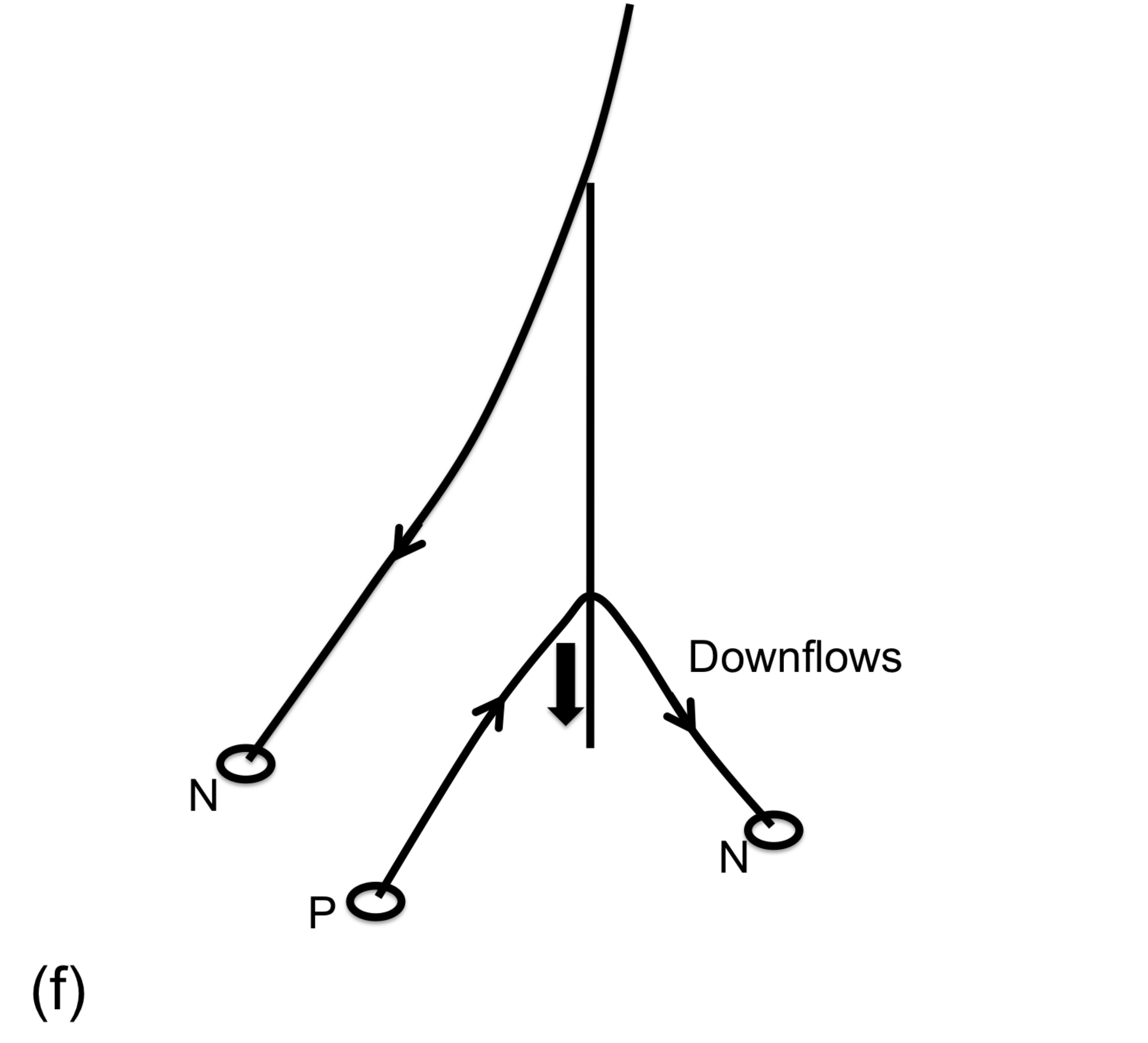}
}

\caption{Cartoon depictions of hypothetical scenarios in which complete loops may contribute as reconnection inflows.}
\label{loop_inflows}
\end{center}
\end{figure*}

It is possible that the flows approaching the candidate current sheet from the south are not involved with the reconnection process, considering their morphology is more like that of complete, stretched loops rather than long field lines anchored within the erupting flux rope.  (A sequence of still images does not capture this morphology satisfactorily, so the reader is referred to the available online movie -- particularly the 171~$\mbox{\AA}$ panel.)  In fact, at least one of these southern loops does appear to be deflected from the reconnection region, although the others are not readily observed again once they reach the candidate current sheet.  Several of these inflowing stretched loops, along with other less discernible inflowing motion, appear from the southern region following the passage of the erupting flux rope.   This inflowing loop motion is also evident in an event occurring on 2011 March 8 \citep{suEA_inprep}.  Figure~\ref{loop_inflows} provides cartoon portrayals of hypothetical scenarios in which complete loops may contribute as reconnection inflows.    

The top panel of Figure~\ref{loop_inflows} considers the merger of two pre-existing loops:  

a)  A region of low magnetic pressure forms in the wake of the erupting flux rope.  Pre-existing loops alongside the axis of the flux rope are swept into this region due to the magnetic pressure imbalance.  The legs of the loops nearest to the low pressure region move faster than those with footpoints further away.   As the flux rope continues to travel outwards, the loops become distorted and possibly kinked as they flow into the evacuated region.  

b)  If the loop legs along either side of the eruption's axis are configured in such a way so as to bring oppositely-polarized field together, reconnection may occur in a very localized region (inset) as it would appear equivalent to Figure~\ref{reconnection} (c).  Otherwise, the loops would be deflected from the current sheet layer.  

c)  Downflowing loops at different heights and orientations result from the reconnection which may require further reconfiguration in order for the field to relax to a potential state.  Standard model upflows extending to the ejected flux rope would not result from this hypothetical reconfiguration since all of the reconnected loops are rooted in the photosphere instead of with the flux rope.  However, relatively short-lived upflow movement may occur near the reconnection sight due to the outflow jets.  The resulting kinked loop would then relax to its potential configuration and would thus appear as a typical downflowing loop for the duration of its movement.

The bottom panel of Figure~\ref{loop_inflows} considers the merger of a pre-existing loop with one polarity of a stretched field line extending to the erupting flux rope.  As in the top panel, the loop and field line are swept into the wake of the erupting flux rope with the possibility of reconnecting through the current sheet.  A downflowing loop and a newly-rooted stretched field line result from the reconnection.  In Section~\ref{inflow101103:orientation}, we will show that the scenario depicted by the bottom panel may be occurring in this flare.  Note that a guide field (into or out of the plane) is likely needed to enable these inherently 3-D scenarios.

%\clearpage

%\clearpage

\subsection{\label{inflow101103:orientation}Field Orientation}

\subsubsection{\label{inflow101103:orientation_feature_matching}Feature Matching}

The SECCHI 195~$\mbox{\AA}$ sequence in Figure~\ref{stereo_seq} shows that the evolution of this flaring region is more complicated than the basic scenario portrayed in Figure~\ref{cartoon_all}, which was assumed to have simpler magnetic topologies.  In particular, the erupting flux rope orientation (east-west) does not appear to be as well aligned with the overall orientation of the developing arcade of interest (southeast-northwest).  In fact, the areas of post reconnection loop development vary considerably in localization and orientation throughout the event.  To properly understand the orientation of the AIA inflow-outflow region, we must identify the corresponding region using \textit{STEREO-B} since the footpoints are obscured by the limb in the AIA field of view.  We match features between the fields of view by comparing SECCHI 195~$\mbox{\AA}$ with AIA 131 \& 193~$\mbox{\AA}$ image sets for $\sim$2.5 days as the active region rotated around the solar limb.  (A movie depicting these matching regions is available via the online supplemental material and is described in the Appendix.)

\begin{figure*}[!ht] 
\begin{center}

\includegraphics[width=.8\textwidth]{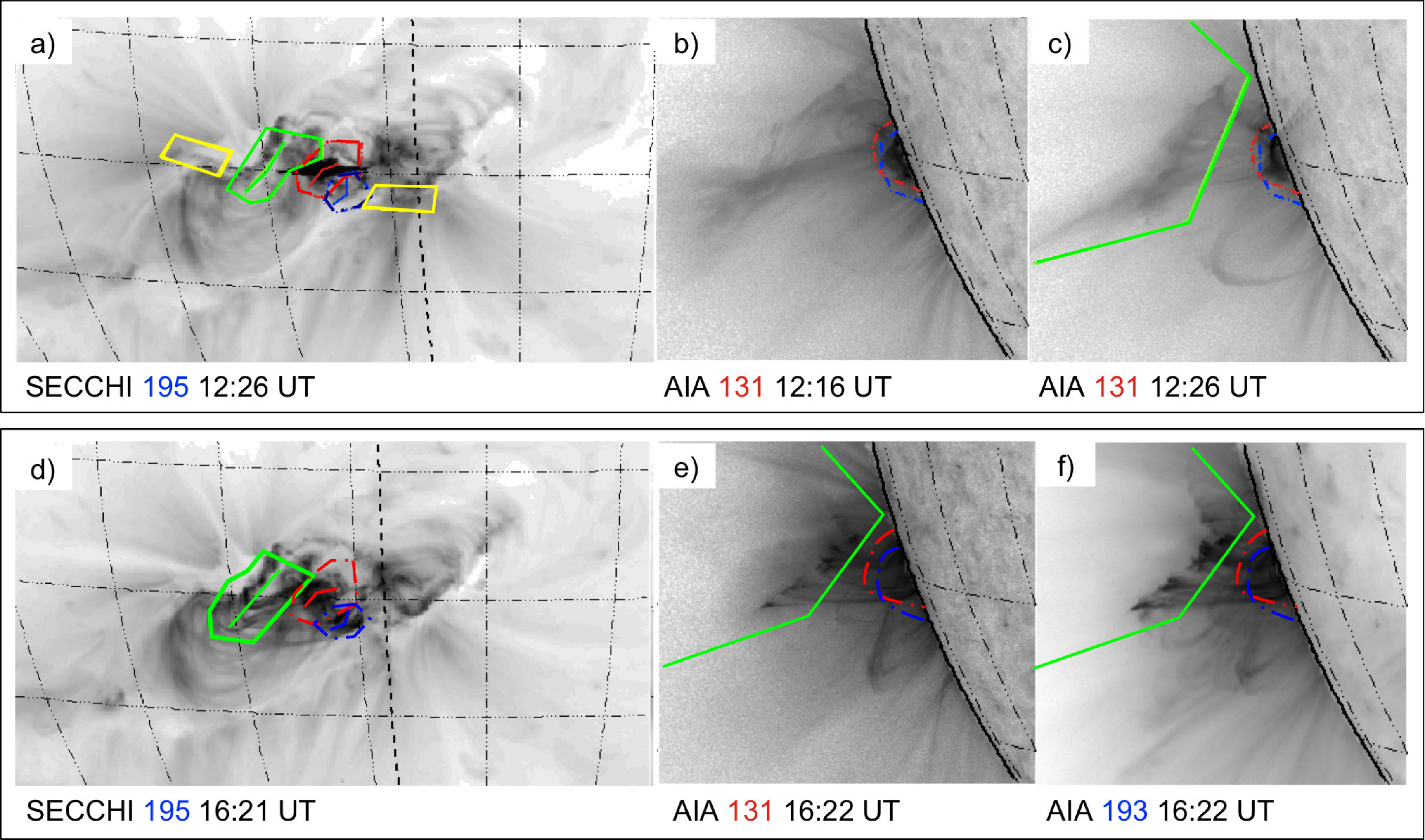}

\caption{Panels a), b), and c) show images taken with SECCHI 195~$\mbox{\AA}$ and AIA 131~$\mbox{\AA}$ near the time of the initial eruption.  The location of the candidate current sheet is noted in panels b) and c) by dashed lines along a bright linear feature.  Images taken with SECCHI 195~$\mbox{\AA}$, AIA 131~$\mbox{\AA}$, and AIA 193~$\mbox{\AA}$ more than four hours into the event are given in panels d), e) and f).  The solar limb as observed from Earth is indicated by the thick dashed longitudinal line in panels a) and d).  The colored regions indicate various features being compared between the \textit{STEREO-B} and \textit{SDO} fields of view (yellow = flux rope footpoints; red/blue = small core flaring regions; green = inflow-outflow region of interest).}
\label{orientation_images}
\end{center}
\end{figure*}

Figure~\ref{orientation_images} provides sample images from the feature matching set.  Panel a) is a SECCHI 195~$\mbox{\AA}$ image taken after the initial eruption as the CME is traversing through the field of view.  The AIA 131~$\mbox{\AA}$ images in panels b) and c) were taken 10 minutes prior to and simultaneously with the SECCHI image, respectively, and correspond to typically hotter temperature plasmas than those observed in panel a).  The footpoint positions of the erupting flux rope are contained within the yellow boxes.  The red and blue boxes indicate the positions of two small core regions of activity that maintain consistent, curved orientations over the course of at least 2.5 days and are the primary flaring sources -- particularly the red region.  Relatively small post-eruption arcades develop within these regions.  De-convolving these arcades within the AIA field of view is complicated by line-of-sight overlap and limb obscuration.  The tops of the corresponding arcades are noted in panels b) and c).  

The inflow-outflow region of interest for this paper is highlighted by the green demarcations in panels a) and c).  During the early phase of the event, this region is not observed using the AIA 193~$\mbox{\AA}$ bandpass.  Because of the lack of emission from this high altitude loop structure at 195 (SECCHI) and 193 (AIA)~$\mbox{\AA}$, the ability to match features is compromised.  While it would appear that this region corresponds to the saturated flaring site within the red box, the orientations are inconsistent.

Comparing images from later in the event, however, allows for more reliable comparisons.  Panel d) is a SECCHI 195~$\mbox{\AA}$ image taken more than 4 hours into the event.  Corresponding images taken with AIA 131 \& 193~$\mbox{\AA}$ are shown in panels e) and f).  At this point in the event, the arcade structure involved with the earlier inflow-outflow observations is also observed in the lower-temperature SECCHI 195~$\mbox{\AA}$ \& AIA 193~$\mbox{\AA}$ passbands.  Because of the similarity in these particular bandpass observations, corresponding features are more readily identified.    The evolution of this event during this later period reveals that the region of interest more likely corresponds to the larger arcade to the east of the core activity.  This southeast-northwest neutral line as seen by \textit{STEREO-B} is consistent with the orientation of the arcade within the AIA field of view and is highlighted by the green demarcations in panels d) through f).  The position of this region has been translated into the early phase of the event and is noted by the green annotations in figures a) and c).  (See the supplemental online movie for further reference.)  

\subsubsection{\label{inflow101103:orientation_mag_pfss}Magnetic Fields}

Because the active region is around the limb at the time of the eruption, direct photospheric magnetic field measurements are unavailable.  We can approximate the field topology by examining magnetograms within the week following the eruption.  Magnetograms taken near the east limb are the most inaccurate due to the lack of magnetic field information prior to crossing the limb; however, magnetograms taken on and after November 5 remain consistent enough to determine approximate locations of opposing magnetic polarities and to derive general topological information through potential field source surface (PFSS) extrapolations \citep{schrijver-derosa_2003}.  PFSS modeling is not expected to accurately portray the fields during eruptions, but it is useful as a tool for visualizing the relaxed field orientation.

\begin{figure*}[!ht] 
\begin{center}

\includegraphics[width=.8\textwidth]{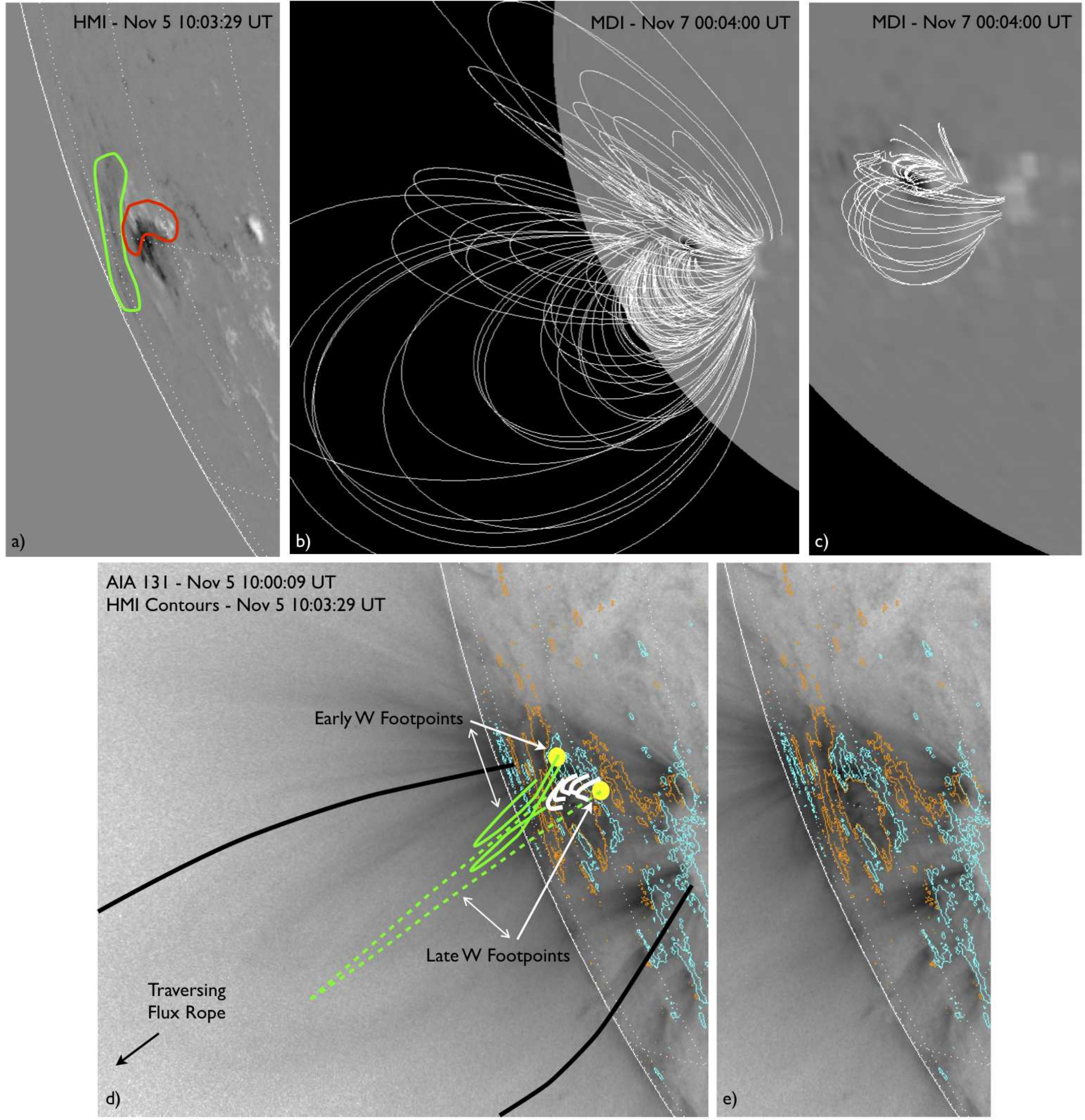}

\caption{Magnetograms and PFSS extrapolations.  a)  HMI magnetogram with the red and green regions from Figure~\ref{orientation_images} demarcated.  b)  MDI magnetogram with PFSS field lines overlaid.  c)  Same PFSS extrapolation as shown in panel b), but with fewer field lines focusing on the core region.  d) \& e)  HMI magnetogram contours (orange/cyan contours refer to negative/positive polarity, respectively) overlaid onto an AIA 131~$\mbox{\AA}$ image.  The approximate orientation of notable field lines are shown.  The white loops correspond to the red core arcade region.  The green field lines correspond to those with eastern footpoints in the green region. The migration of the corresponding western footpoints is noted.}
\label{mag_pfss}
\end{center}
\end{figure*}

Figure~\ref{mag_pfss} provides this magnetic field information.  An \textit{SDO}/HMI magnetogram, from $\sim$2 days after the initial eruption, is shown in a) with the red and green regions from Figure~\ref{orientation_images} demarcated.  A PFSS extrapolation showing the potential magnetic topology for the entire region is shown in b) overlaid onto a \textit{SOHO}/MDI magnetogram, taken nearly 4 days post-eruption.  Note that due to the complexity of the underlying fields, there are several different field line orientations crossing paths at various heights.  Panel c) is the same extrapolation as shown in b) but focusing on the core region with fewer field lines.  

The PFSS model as applied to this flaring region is imprecise because 1) the region is behind the limb, and thus, the magnetic topology is derived from data several days after the initial flare, 2) the pre-eruption active region is certainly not potential considering the ample flare production, and 3) magnetogram data is most unreliable on the east limb.  Therefore, these extrapolations are not expected to exactly map out the field lines observed with SECCHI and AIA; however, they do serve to show that the various overlapping field line orientations are evident even at these relatively low heights and in a confined region.

The contours from the magnetogram in Figure~\ref{mag_pfss} a) are overlaid onto a corresponding AIA 131~$\mbox{\AA}$ image in panels d) and e)  (orange/cyan contours refer to negative/positive polarity, respectively).  The approximate orientation of the field lines relating to the features of interest in Section~\ref{inflow101103:orientation_feature_matching} are shown in panel d).  The white loops correspond to the red core arcade region.  The orientation of the white loops as drawn is based upon comparison between the HMI magnetic field contours and the loops actually observed by \textit{STEREO-B} for at least 2.5 days following the initial eruption.  We attribute the slight difference in loop appearance between those in panels c) and d) in this core region as being due to using a potential extrapolation near the limb, 2 days of additional rotation, and possible shear (a typical characteristic of pre-eruptive regions along polarity inversion lines \citep{su-golub-vanball_2007}). 

The green lines correspond to the development of the field in the green region of interest.  These loops appear kinked and stretched in the SECCHI images and are therefore highly non-potential.  The western footpoints of these loops migrate after several hours possibly due to reconnection higher in the corona where the field lines have different footpoints, as is noted above in the PFSS extrapolations.

The actual flaring site is located within the red region of interest where the first post-eruption arcade develops.  This region also corresponds to the  strongest polarity inversion.  Due to limb obscuration, the original erupting flux rope is not imaged by AIA nor is it readily apparent in the SECCHI observations.  (A long filamentary structure is seen in the SECCHI images in the far south of the region but is not noticeably involved with the eruption.)  Fortuitously, two eruptions originate from this region over the following several days wherein a flux rope is observed to erupt from the red core region.  (Refer to the available online movie.)  

By analyzing the subsequent eruptions and combining this information with the PFSS extrapolations, we infer that a relatively small flux rope erupts from within the core red region.  As it travels into the corona, it encounters the overlying field traversing the entire region in the general east-west direction (Figure~\ref{mag_pfss} b)).  This overlying field combines with the original erupting flux rope (possibly via reconnection or entanglement) to create a larger erupting flux rope which disrupts the field above the entire region.  This process may be likened to the flare breakout model (\citeauthor{macneiceEA_2004} \citeyear{macneiceEA_2004}; \citeauthor{antiochos-devore-klimchuk_1999} \citeyear{antiochos-devore-klimchuk_1999}) although not a necessary scenario due to the uncertainty in the original flux rope and overlying field structure; however, a truly 3-D model addressing the various overlying field line orientations, bend in the core polarity inversion line, and complex photospheric polarity organization is necessary in order understand the evolution of the flux rope.  

The focus of this paper is to discuss the flows occurring in the wake of the flux rope; therefore, we will refer to the erupting flux rope for this region as the large, combined one traversing the region since it is required to create the post-eruption arcade in the green domain -- where the flows of interest are occurring.  (The post-eruption arcade within the core red region could develop independently due to the original smaller flux rope eruption.)

\subsubsection{\label{inflow101103:orientation_topology}Extrapolated Topology}

\begin{figure*}[!ht] 
\begin{center}

\includegraphics[width=.9\textwidth]{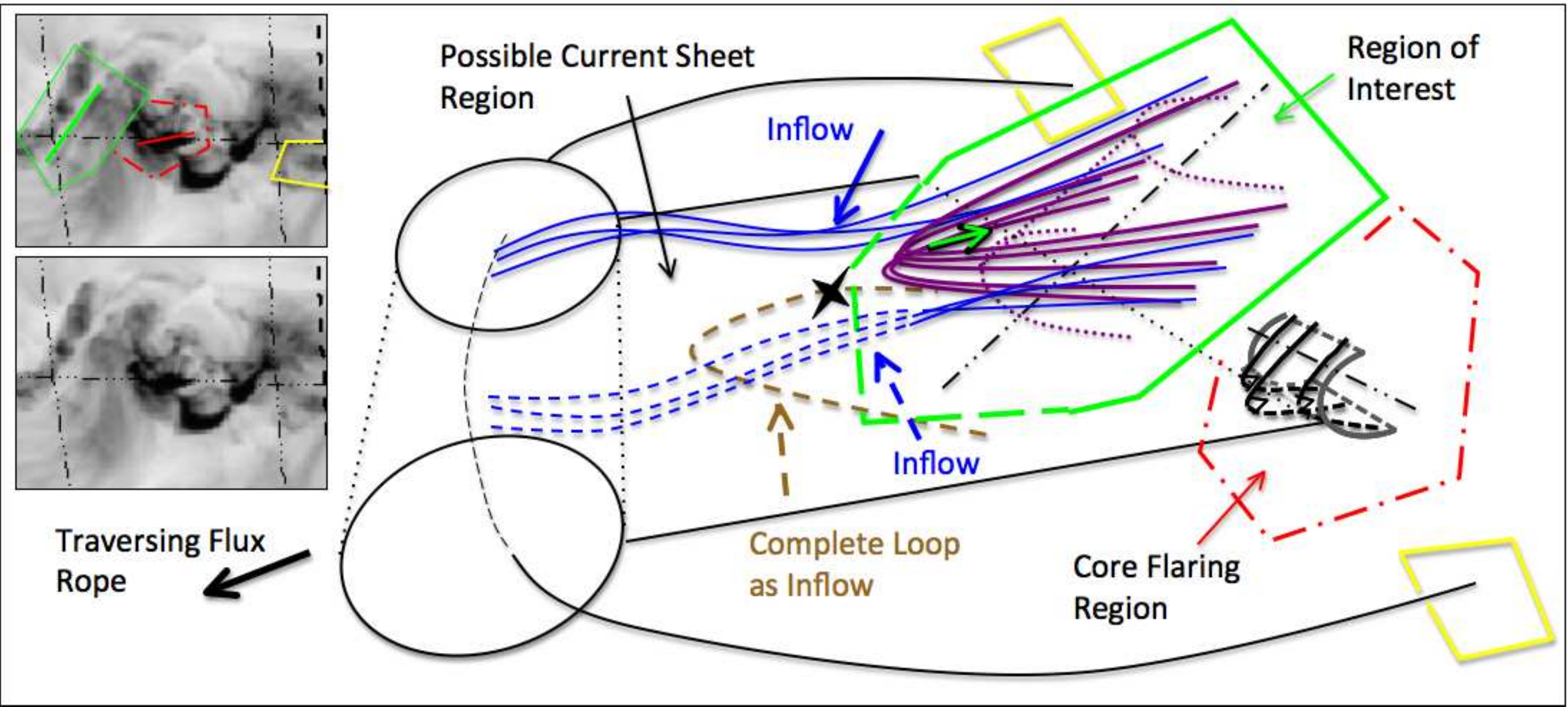}

\caption{Simplified diagram of the general orientation of the features depicted in Figure~\ref{orientation_images}.  The enlarged, cropped SECCHI 195~$\mbox{\AA}$ inset images were taken at 12:41~UT and are provided for comparison to the diagram.}
\label{orientation_cartoon}
\end{center}
\end{figure*}

Using these image sets and the magnetic field topological information as a guide combined with the basic reconnection scenario depicted in Figure~\ref{cartoon_all}, a simplified description of the orientations of the most relevant features (i.e., traversing erupting flux rope as discussed above in Section~\ref{inflow101103:orientation_mag_pfss}, inflow-outflow region (green), region where a current sheet may potentially develop, and core flaring site (red)) is proposed in Figure~\ref{orientation_cartoon}.  (To reduce clutter in the diagram, the blue core flaring site is omitted.)  The inset contains two duplicate images taken with SECCHI 195~$\mbox{\AA}$ at 12:41~UT.  The images have been enlarged and cropped in order to focus on the regions relevant to the diagram at right.  The top image of the inset has been annotated as a guide for comparing the inflow-outflow region of interest (green), the core flaring region (red), and one of the flux rope footpoints (yellow) to the diagram.  For the red core region, we are focusing on the consistently emitting east-west arcade; however, the lower left portion of the arcade appears to curve to the south during some phases of the event.

In this interpretation, the combined flux rope traverses across the entire region and intersects the polarity inversion line (PIL) of the inflow-outflow region at nearly a perpendicular angle.  As the flux rope erupts, it drags, stretches, and twists the field lines (blue) that cross this PIL.  These field lines become inflows as they approach the current sheet created in the wake of the flux rope.  As discussed above in Section~\ref{inflow101103:inflows}, the field lines below the current sheet may be the pre-existing, large complete loops to the south of the region of interest.  These loops being swept toward the current sheet could possibly behave as inflows as indicated by the dashed brown field line (see also Figure~\ref{loop_inflows} - bottom panel).  

The reconnection region is limited to the intersection position with the current sheet.  The newly-reconnected, highly stretched and twisted field lines eventually relax to form a typical post-eruption arcade (dotted purple).  So while the flux rope, which is oriented in the east-west direction, dictates the north-south orientation of the inflows, the arcade is structured according to the footpoints of the field lines crossing the PIL in the southeast-northwest direction.  The twisting of this region thereby allows for the unusual angle between the inflows and outflows described herein (refer to Figures~\ref{tracks}~\&~\ref{stackplot_outflows_131}).

\subsection{\label{inflow101103:inflow_measurements}Inflow Measurements}

\begin{figure*}[!ht] 
\begin{center}

\includegraphics[width=.85\textwidth]{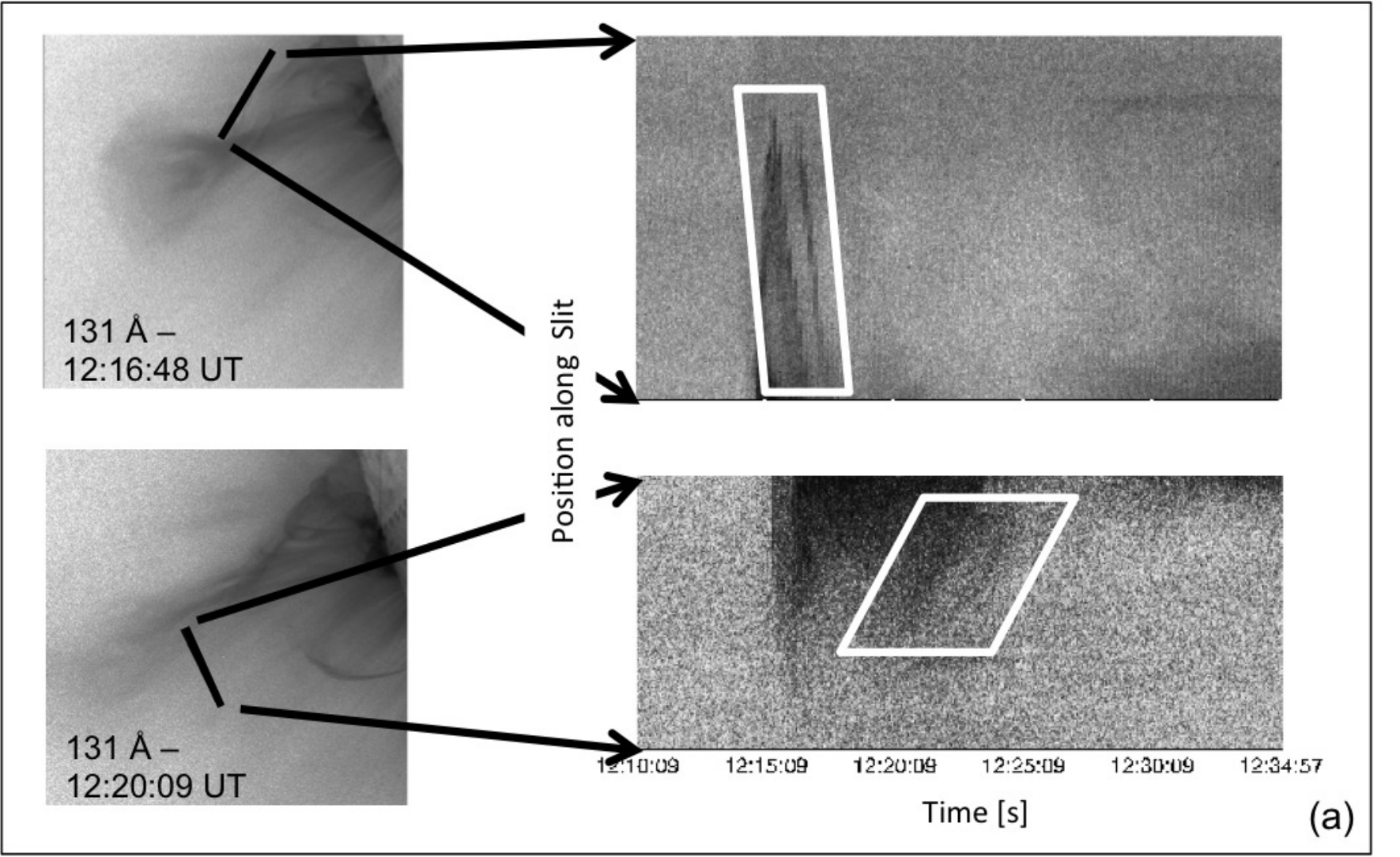}
\includegraphics[width=.85\textwidth]{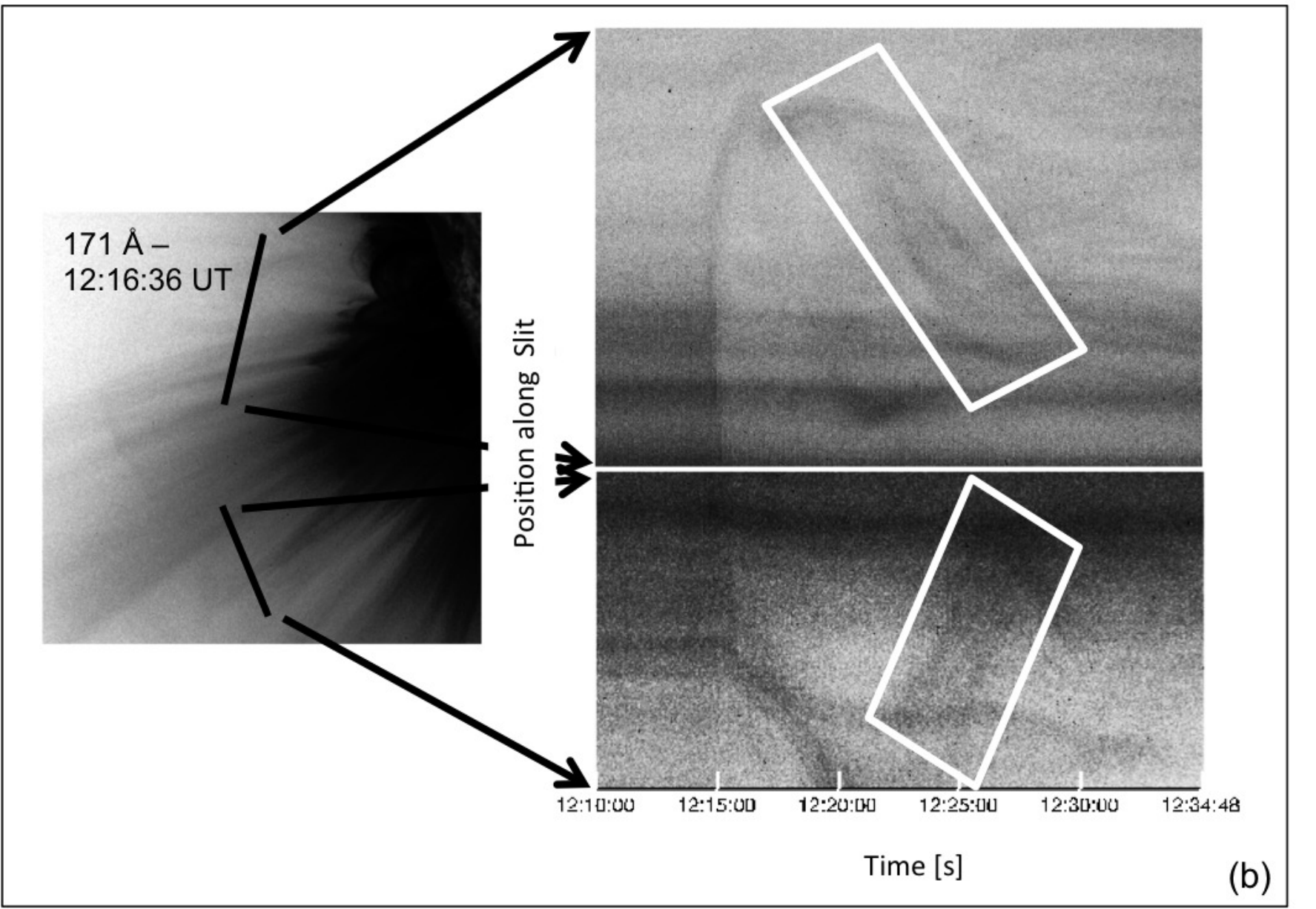}

\caption{Position versus time stackplots made by extracting three-pixel wide slits (depicted as black lines in each panel on the left) from an (a) AIA 131 $\mbox{\AA}$  and (b) AIA 171 $\mbox{\AA}$ image sequence.  Each three-pixel segment represents $\sim$12 seconds between 12:10 and 12:38~UT.  The arrows indicate the orientation of the slits with respect to the corresponding stackplots.  The boxed regions of the stackplots highlight the inflows into the candidate current sheet.}
\label{stackplot_inflows_131_171}
\end{center}
\end{figure*}

Figure~\ref{stackplot_inflows_131_171} contains position-versus-time stackplots, which we constructed by extracting three-pixel wide cross-sections from each image file.  (Stackplots are only provided for the 131 and 171~$\mbox{\AA}$ channels because they are the most discernible and representative.)  In these stackplots, each individual cross section in the plot represents 12 seconds---the AIA observation cadence during the eruption.  We have indicated the location of each cross-section in a single frame of the AIA movie shown to the left of each stackplot in the figure. White boxes in the stackplots highlight the inflows into the candidate current sheet. Inflows directed into the candidate current sheet from the north are more prominent than those from the south.  Projection, line-of-sight effects, and twist of the region (as discussed above) may add to the observational discrepancies between the northern and southern inflows. 

%  Although \textit{STEREO~B's} vantage point directly above the flare provides 3-dimensional context to the region, the inflowing features seen with AIA do not obviously correspond with any SECCHI features, likely due to the low image cadence ($\sim$5 minutes) and spatial resolution (\textgreater 2.5 times poorer than AIA).

\begin{figure*}[!ht] 
\begin{center}

\includegraphics[width=1\textwidth]{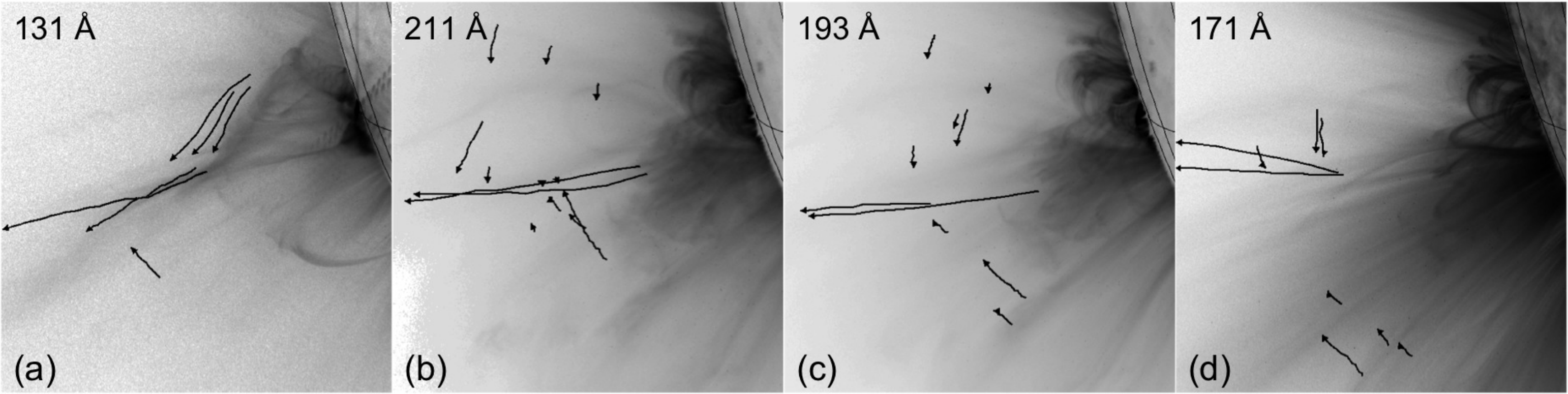}

\caption{flux rope and inflow tracks from AIA (a) 131~$\mbox{\AA}$, (b) 211~$\mbox{\AA}$, (c) 193~$\mbox{\AA}$, and (d) 171~$\mbox{\AA}$.  The tracks moving outward away from the solar surface are associated with the erupting flux rope instead of the inflows.}
\label{tracks}
\end{center}
\end{figure*}

To determine the inflow velocities, we manually tracked the portion of the flow that appears to travel directly into the apparent reconnection region.  Because the active region is only slightly behind the limb ($\sim$0 -- 10$^{\circ}$), the errors in de-projecting the position measurements would be overwhelmed by the complicated topology of the coronal magnetic field structure (as described in Section~\ref{inflow101103:orientation}); therefore, we report only plane-of-sky measurements for positions and speeds. The resulting tracks are overlaid onto a corresponding wavelength image in Figure~\ref{tracks}.  Tracks moving outward away from the solar surface are associated with portions of the erupting flux rope (i.e. initial fast-moving edge [211, 193, \& 171~$\mbox{\AA}$], primary leading edge [131, 211, 193, \& 171~$\mbox{\AA}$], and trailing edge [131~$\mbox{\AA}$]).  The high-temperature flow trajectories are more inclined into the candidate current sheet than those at low temperatures as is evident when comparing Figure~\ref{tracks} (a) to the other three panels. 

\begin{figure*}[!ht] 
\begin{center}

\includegraphics[height=125pt]{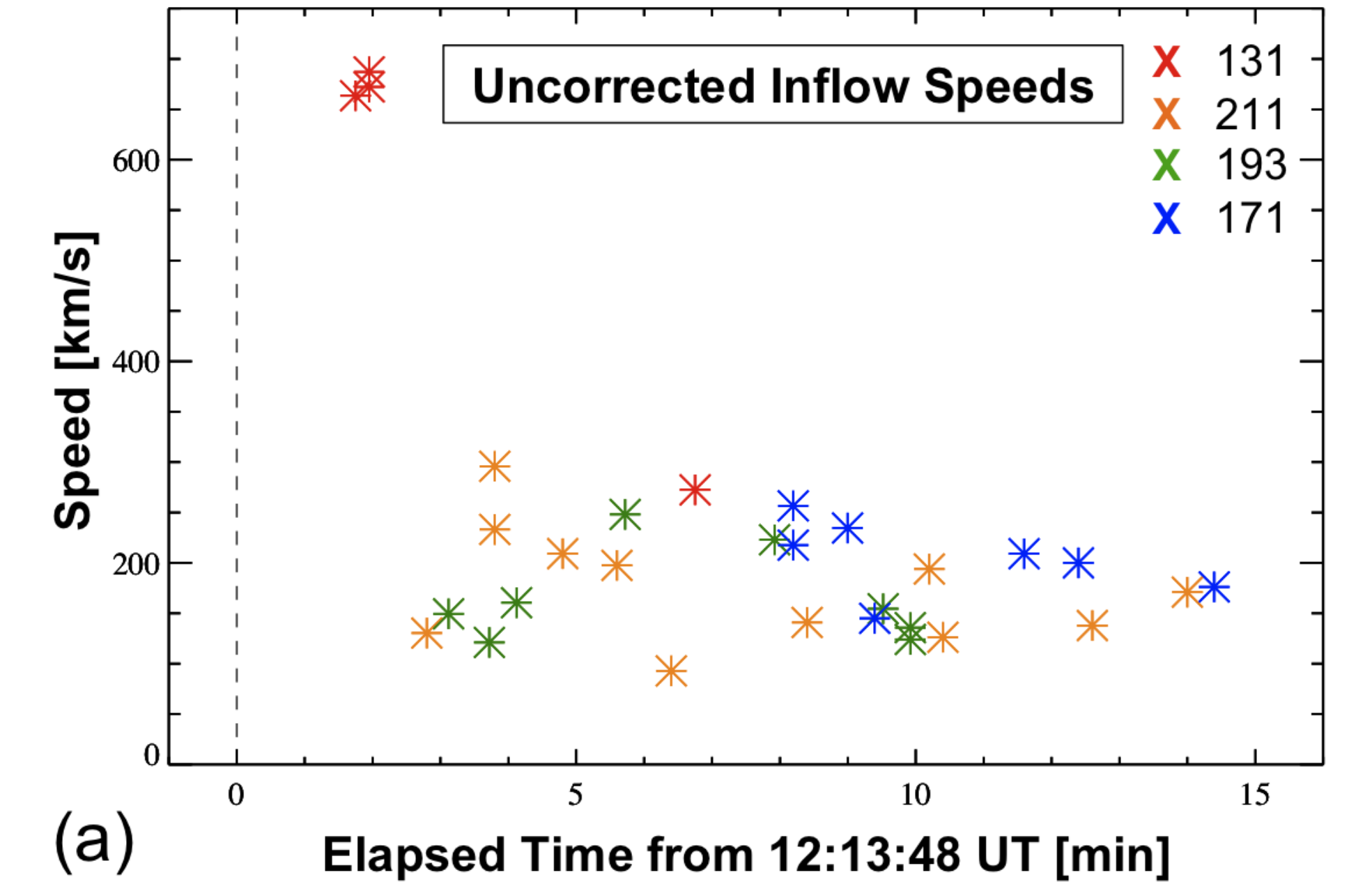}
\includegraphics[height=125pt]{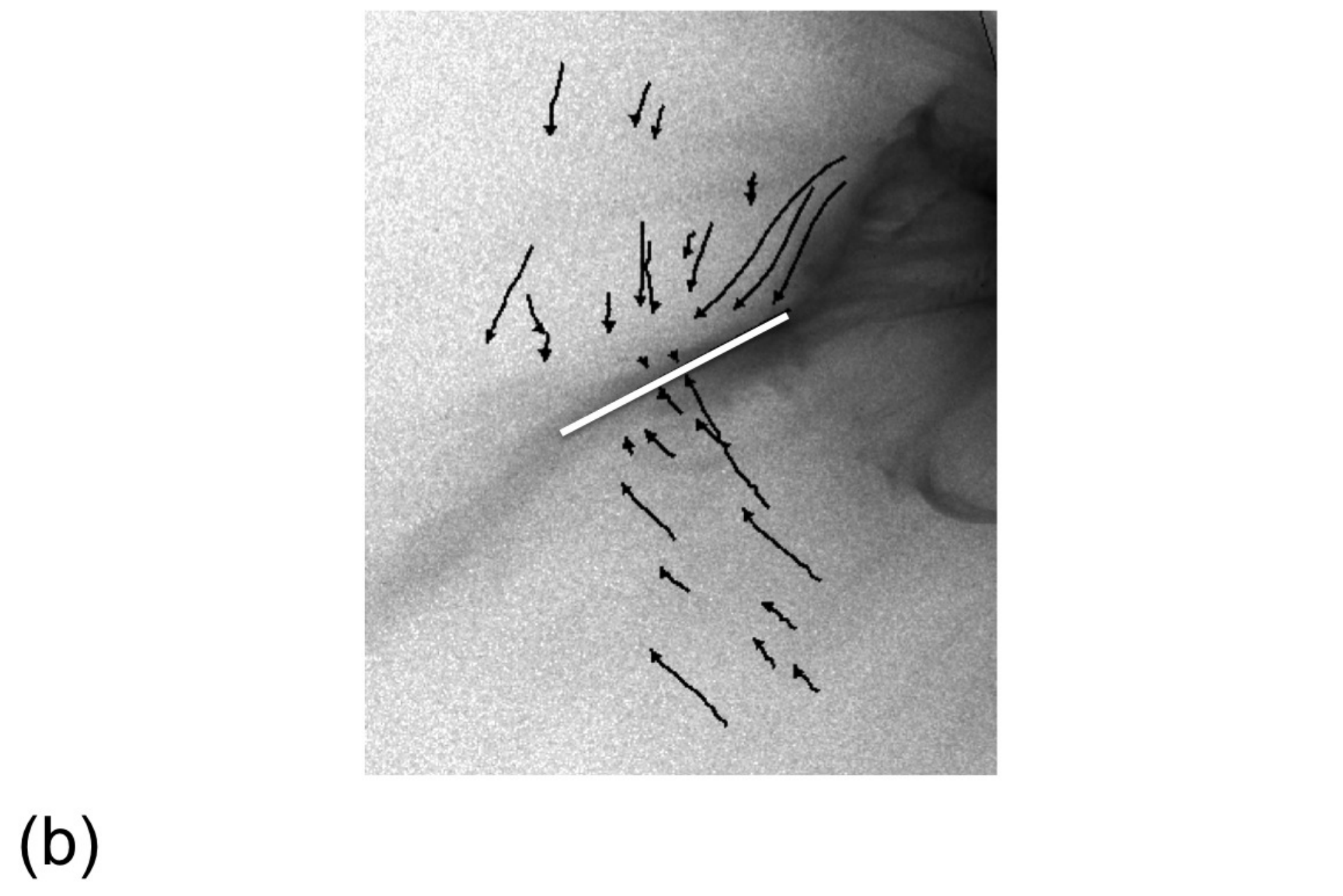}
\includegraphics[height=125pt]{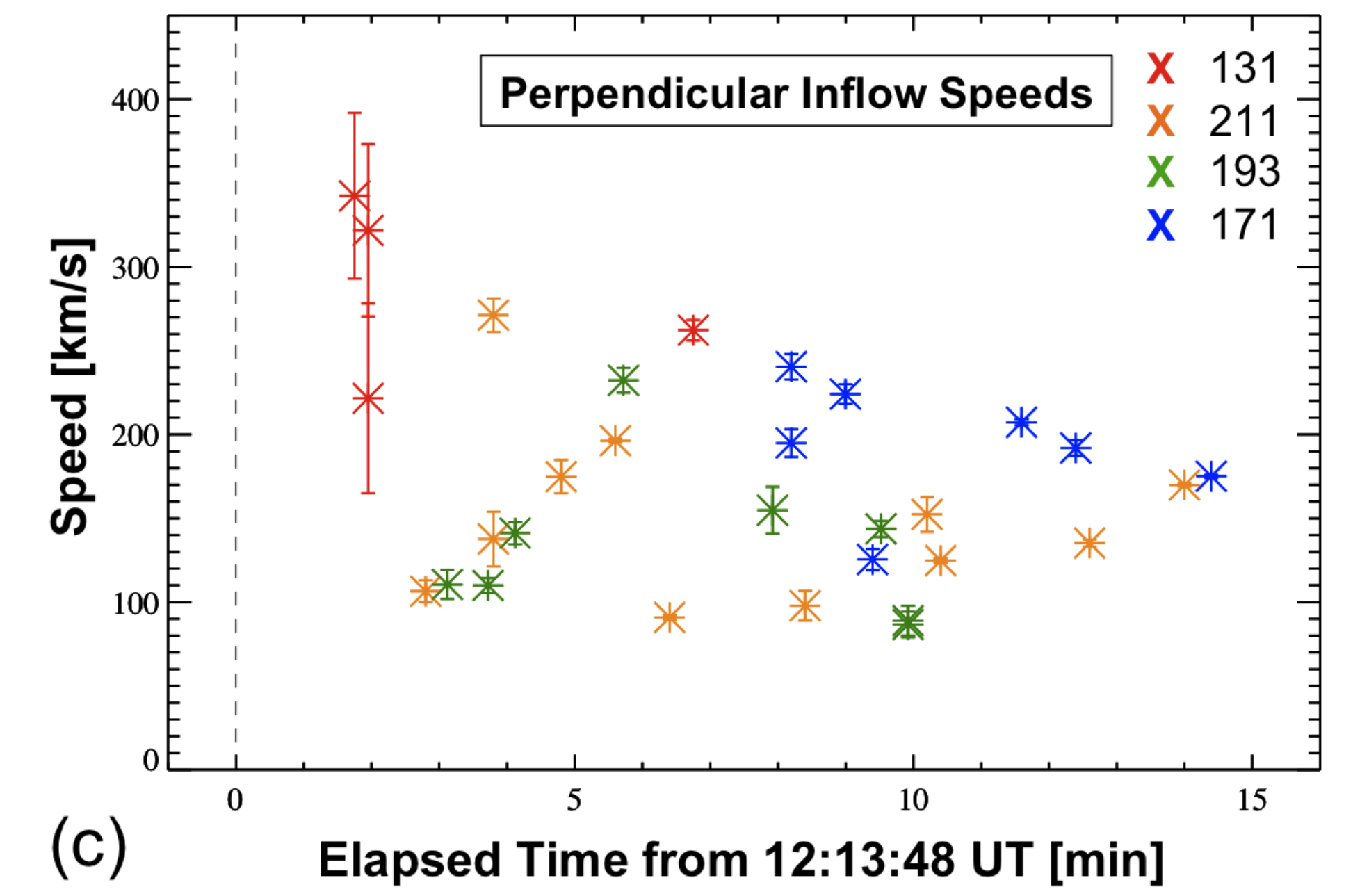}
\includegraphics[height=125pt]{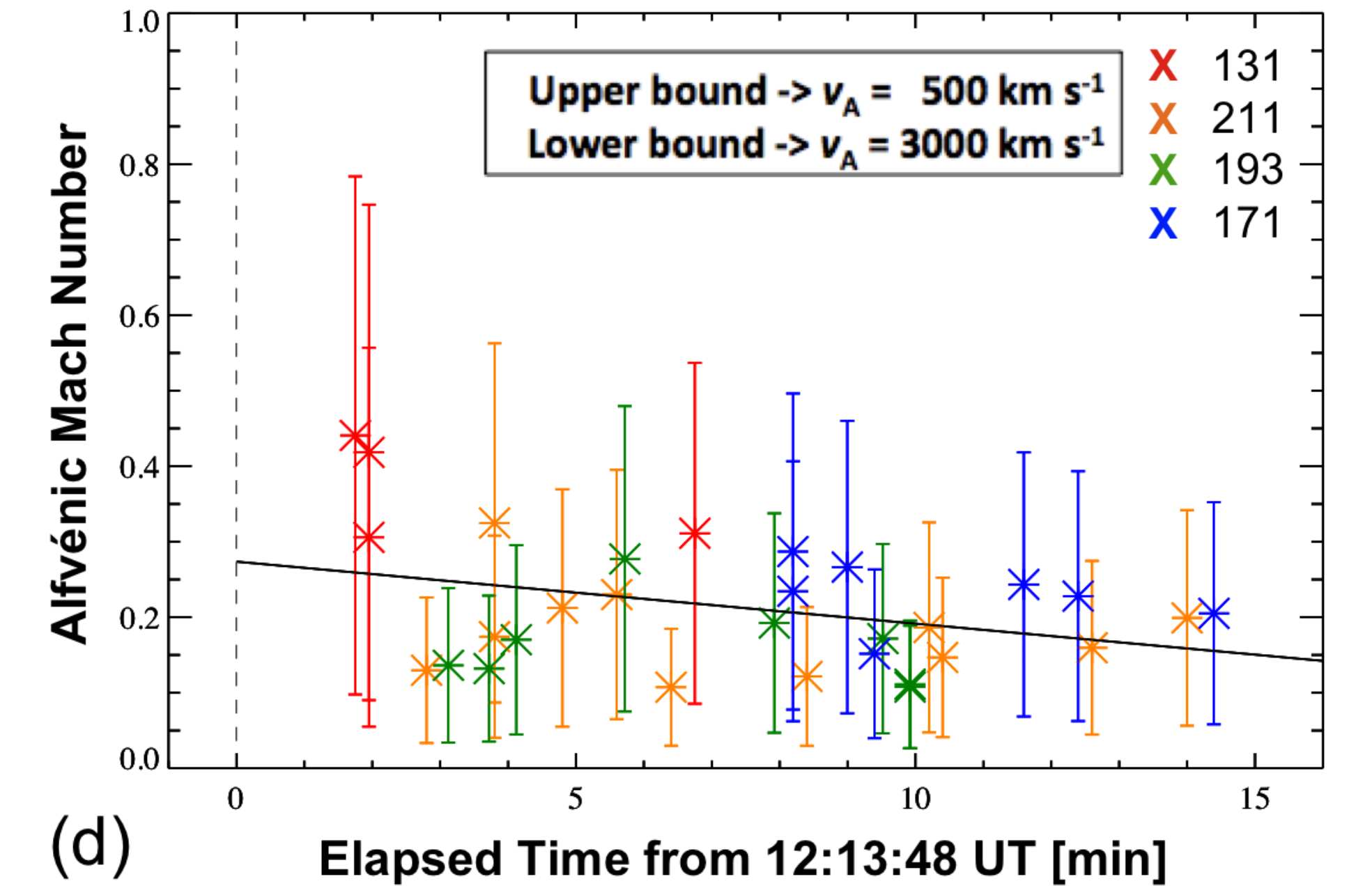}

\caption{(a)  Estimated plane-of-sky inflow speeds versus elapsed time from the start of the first flux rope track (12:13:48 UT).  (b)  Composite image showing the inflow tracks from all 4 selected AIA filters overlaid onto a 131~$\mbox{\AA}$ image (12:23:45~UT).  The white line indicates the candidate current sheet position used to determine the perpendicular component of the inflow speeds with respect to the current sheet ($\pm~5^{\circ}$).  (c)  Estimated plane-of-sky perpendicular inflow speeds versus elapsed time.  (d)  Estimated Alfv\'{e}nic mach number (M$_{A}$) versus elapsed time.  The bars represent the range of M$_{A}$ values obtained by varying the Alfv\'{e}n speed from 500 to 3000~km~s$^{-1}$ with the asterisk positioned at the mean value.  A linear fit is overlaid onto (d) to point out an apparent decreasing trend of M$_{A}$ with time (as long as the 131~$\mbox{\AA}$ speeds are included).}
\label{speeds_ma}
\end{center}
\end{figure*}

We use an uncertainty of $\pm$~3 pixels ($\sim$~2~arcsecs) to account for errors in manually assigning positions, but we note that the resulting uncertainty bars for the speeds we determine are smaller than the symbols in Figure~\ref{speeds_ma} (a).  However, these uncertainties do not include the error associated with attempting to track a coherent structure or the original choice in position along the feature which is tracked.  Such uncertainties are difficult to quantify and rely on several assumptions, so the speeds provided should be considered moderately imprecise.  

Bearing the uncertainties in mind, an overall decrease in average speed with respect to elapsed time can be inferred from the figure and is clearly corroborated by the image sequences.  (Refer to the available online movie.)  It should be noted, however, that without the 131~$\mbox{\AA}$ speeds, the decreasing trend is less pronounced.  The speeds from the other wavelength measurements could agree with a relatively constant speed with time.  The high temperature 131~$\mbox{\AA}$ inflows primarily occur early in the eruption immediately in the wake of the flux rope ($\sim$12:15 -- 12:18 UT).  The first three of these inflows are also the fastest, with estimated velocities between 660 and 690 km s$^{-1}$.  Inflows do not become apparent in the generally lowest temperature 171 $\mbox{\AA}$ channel until $\sim$12:22 UT and are slower ($\sim$150 $-$ 260 km s$^{-1}$) (see Figure~\ref{speeds_ma} (a) for a depiction of all inflow speeds versus time per filter).

The perpendicular component of the inflow speed with respect to the current sheet is more relevant to the reconnection process; therefore, these average perpendicular speeds were calculated and provided in Figure~\ref{speeds_ma} (c) based on the position of the candidate current sheet (highlighted by the white line overlaid onto the composite track image in panel (b)).  The error bars result from applying an uncertainty of 5$^{\circ}$ in the perpendicular angle.  The decrease in average speed with time remains apparent.  The inflow velocities reported by \citeauthor{narukage-shibata_2006} (\citeyear{narukage-shibata_2006}) are given as 2.5 times the apparent inward motion based on \citeauthor{chenEA_2004} (\citeyear{chenEA_2004})'s numerical simulations.  The rising motion of the reconnection X-point is also taken into consideration by these authors as well as \citeauthor{yokoyamaEA_2001} (\citeyear{yokoyamaEA_2001}), where $v_{inflow}~=~v_{inward}~-~v_{xp}$.  In this paper, however, we only report the inflow velocities as the perpendicular apparent inward motion ($v_{inflow}~\sim~v_{\perp, inward}$) because the limb-obscuration makes X-point motion unclear and the speed uncertainties are inherently large.

As described in \citet{isobe-takasaki-shibata_2005} and \citet{yokoyamaEA_2001}, we attempt to determine Alfv\'{e}nic Mach numbers (M$_{A}$) for each of the inflow speeds using M$_{A} = v_{in}/v_{A}$, where $v_{in}$ is the inflow speed and $v_{A}$ is the Alfv\'{e}n speed upstream of the reconnection site.  Instead of folding several assumptions into an estimate for $v_{A}$ (which cannot be directly measured with current technology), we provide a range of possible M$_{A}$ values using a reasonable range of coronal Alfv\'{e}n speeds.  The range of speeds used (500 $-$~3000~km~s$^{-1}$) encompasses the typically assumed coronal Alfv\'{e}n speed of 1000~km~s$^{-1}$ with allowances for deviations due to magnetic field and density fluctuations.  \citeauthor{narukage-shibata_2006} (\citeyear{narukage-shibata_2006}) derived a coronal field strength by comparing the speed of footpoint separation to that of the inflows.  This method is not available for this flare as the footpoints are behind the limb.

The derived values of M$_{A}$ are plotted with respect to elapsed time in Figure~\ref{speeds_ma}~(d).  The decreasing trend mimics that of the inflow speeds shown in panel~(c).  The average of the range of M$_{A}$ values for this early time period decreases from $\sim$0.4 -- 0.1.  We note that while these rates are systematically overestimated, even within the large uncertainties, these high values are more appropriate to fast, Petschek-type reconnection versus the slower Sweet-Parker reconnection models, at least during this phase of the eruption.

\clearpage

\clearpage

\subsection{\label{inflows101103:outflows}Reconnection Outflows}

Between 12:10 and 12:38~UT, we can clearly track several supra-arcade downflowing loops (SADLs) that flow along the candidate current sheet in the high-temperature 131~$\mbox{\AA}$ bandpass.  (This SADL movement can also be seen in the 94~$\mbox{\AA}$ bandpass, but with less clarity.)  The 131~$\mbox{\AA}$ flows later in this time period (after $\sim$12:32~UT) moving in from beyond the field of view look more similar to SADs or plasmoids than SADLs, possibly due to the level of noise at those heights.   

\begin{figure*}[!ht] 
\begin{center}

\includegraphics[width=.9\textwidth]{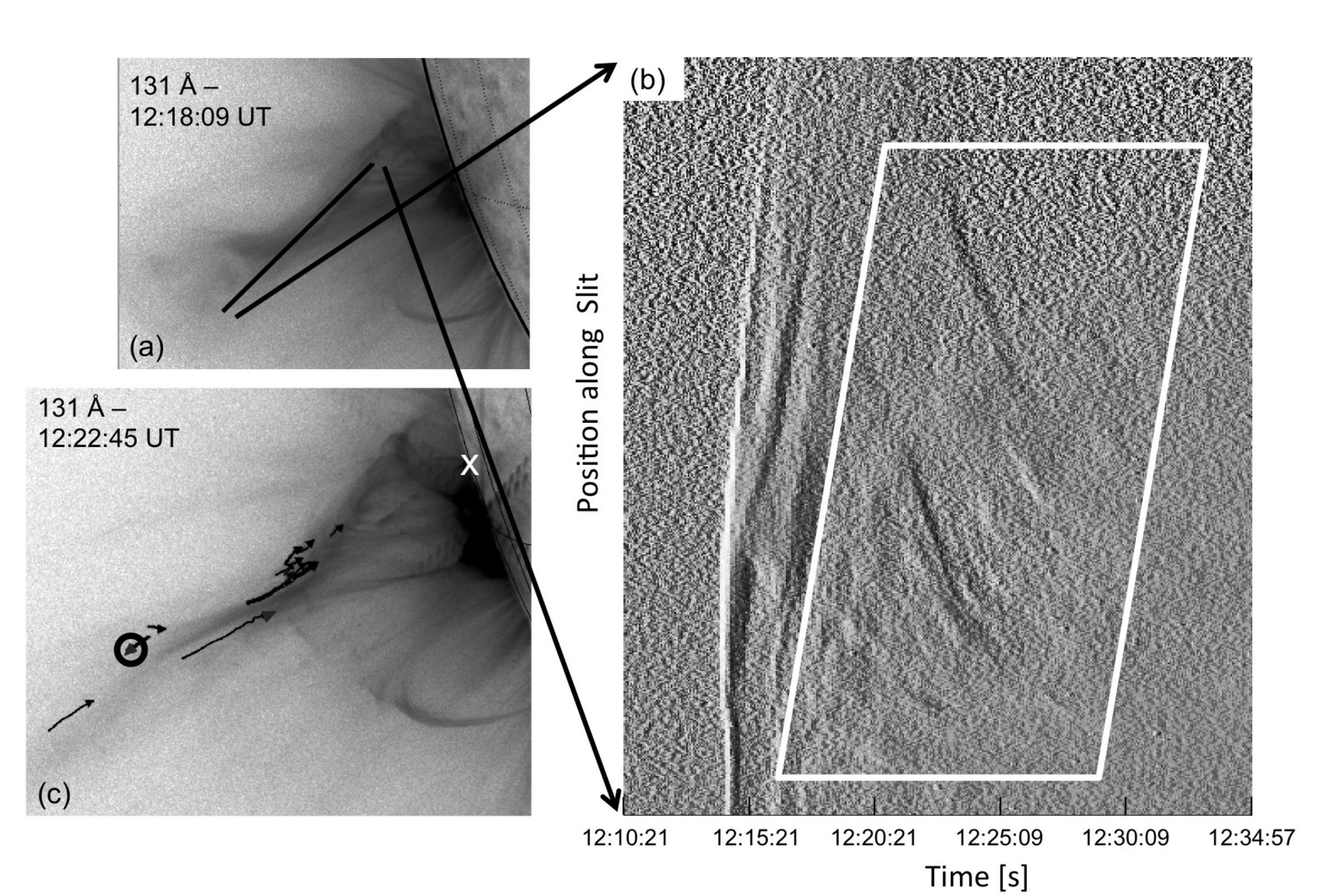}

\caption{Position versus time stackplot (b) made by extracting three-pixel wide slits (depicted as a black line in the reverse-scale image at left (a)) from an AIA 131 $\mbox{\AA}$ image sequence.  Each three-pixel segment represents $\sim$12 seconds between 12:10 and 12:38~UT.  The arrows between the images indicate the orientation of the slit with respect to the stackplot.  The boxed region of the stackplot highlights the outflows along the candidate current sheet.  The downflow tracks, with an upflow track highlighted by a circle, are shown in (c).  The white ``X" marks the footpoint used to determine outflow heights (90E, 18S).}
\label{stackplot_outflows_131}
\end{center}
\end{figure*}

Reconnection inflows and outflows continue beyond this focused time period in several AIA channels.  Figure~\ref{stackplot_outflows_131} provides a stackplot similar to that seen in Figure~\ref{stackplot_inflows_131_171}, with the corresponding image cross-section focused on the downflows in 131~$\mbox{\AA}$.  The 12 second-cadence image set used to create the stackplot was run-differenced and contrast-enhanced using histogram equalization.  The downflows appear as descending dark and bright lanes inside the white box overlaid on the plot.  (Due to the enhancement process which manipulates the emission values, the stackplot is intended only to provide a visual reference of the downflowing tracks rather than to make a statement concerning the brightness of the loops.)

\begin{figure*}[!ht] 
\begin{center}

\includegraphics[height=145pt]{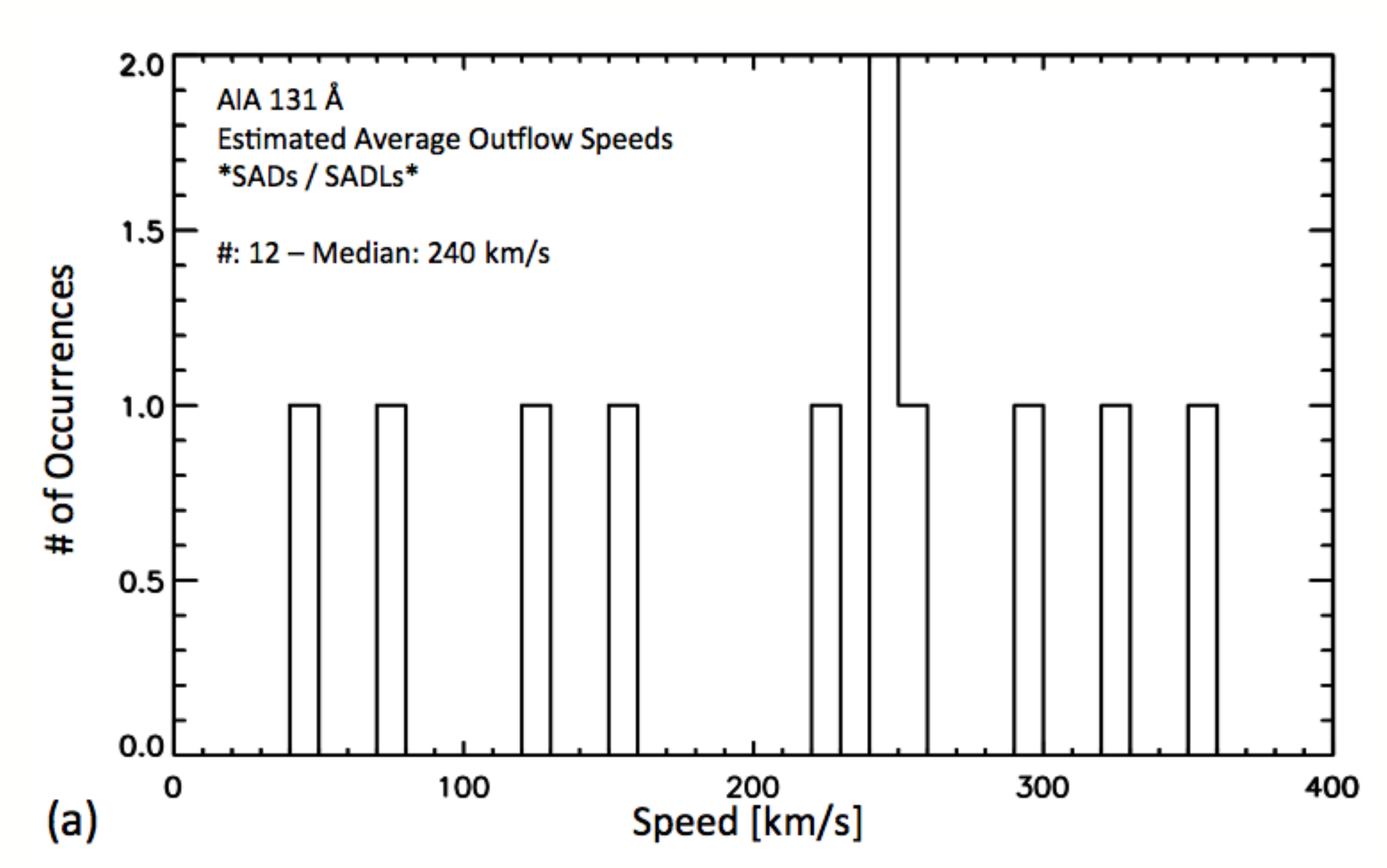}
\includegraphics[height=145pt]{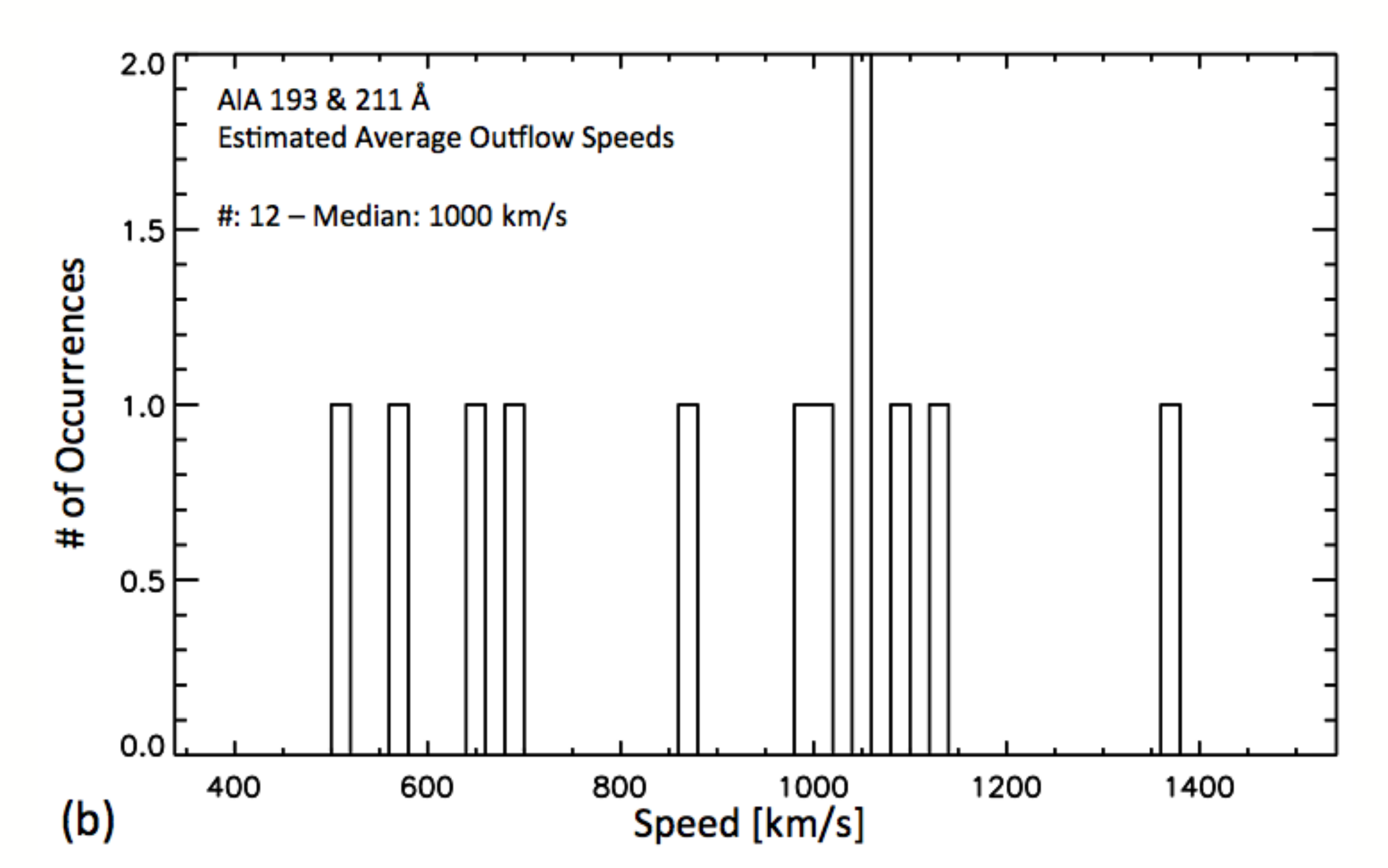}
\includegraphics[height=145pt]{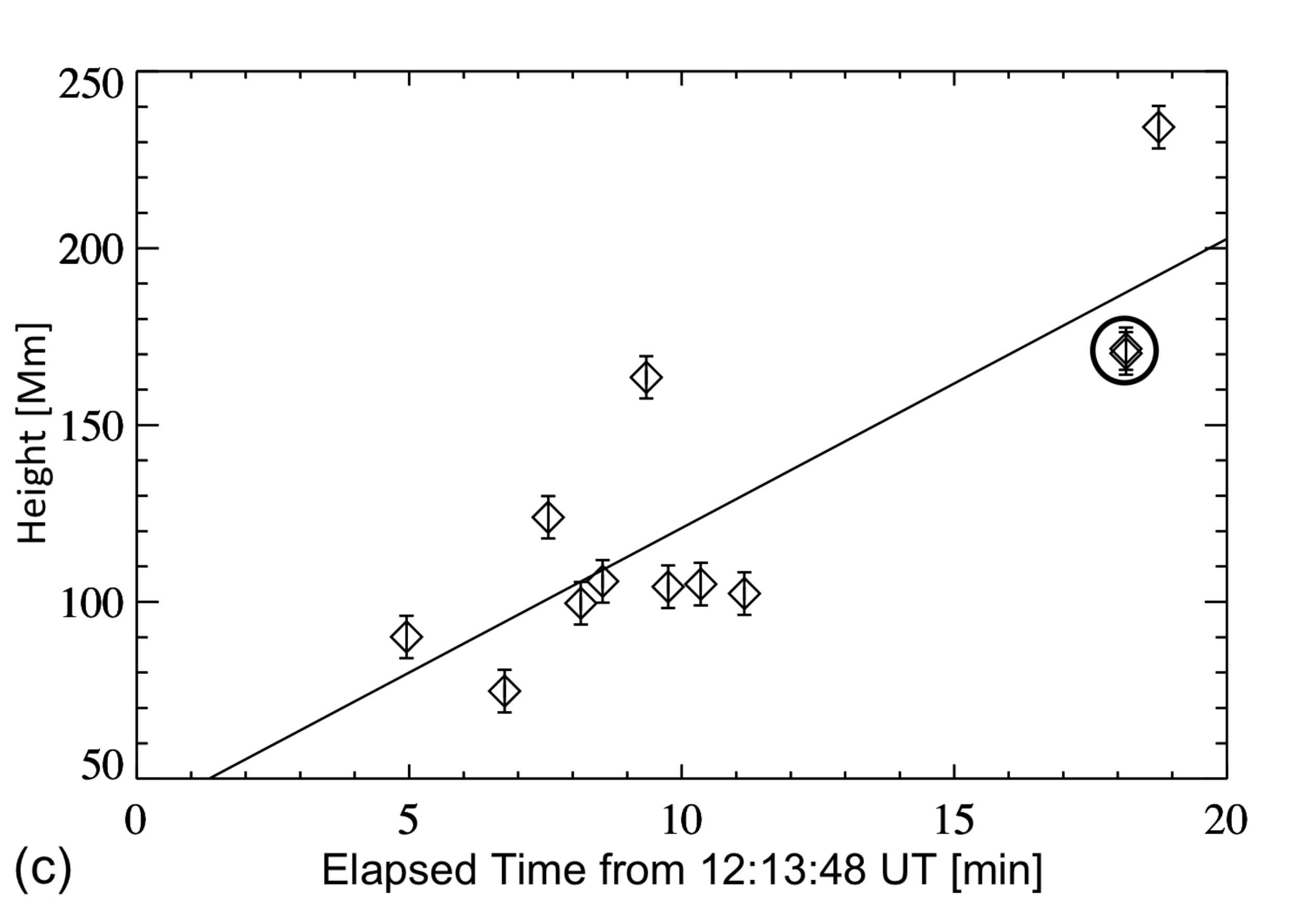}

\caption{(a)  Histogram of average outflow speeds measured for the SADs and SADLs tracked in AIA 131.  (b)  Histogram of average outflow speeds measured for the flows in AIA 211 and 193.  These flows do not have the same morphology as typical SADs and SADLs, but instead appear as material moving down the legs of the newly-reconnected loops.  (c)  Estimated initial heights versus time plot for the outflows tracked in AIA 131.  The linear fit suggests an upward progression of initial height \textit{detections}.  A possible disconnection event, wherein an upflow and downflow appear to originate from nearly the same region simultaneously, is indicated by the circle.}
\label{outflow_speeds_heights}
\end{center}
\end{figure*}

Sunward flowing material can be seen in the  211 and 193~$\mbox{\AA}$ bandpasses, but not in the typical SAD or SADL morphology.  Instead, these flows appear as density enhancements traveling down along the legs of newly-formed loops which could be interpreted as moving material, shocks (\citeauthor{guidoni-longcope_2011b} \citeyear{guidoni-longcope_2011b}; \citeauthor{guidoni-longcope_2011} \citeyear{guidoni-longcope_2011}), or waves, possibly fast mode.  When these flows are overlaid onto the AIA 131 image sequence, they do indeed move along the legs associated with the SADLs, but at speeds an order of magnitude faster than downward speeds of the apexes of the loops (Figure~\ref{outflow_speeds_heights}).  Outflows are not apparent in the 171 channel.

Because it is difficult to separate fine structure and acquire reliable measurements using stackplots, we again tracked the downflowing loops manually (as indicated in the online movie) in order to estimate flow times, velocities, and heights.  Figure~\ref{outflow_speeds_heights} provides estimated average plane-of-sky outflow speeds for (a) the SADs/SADLs tracked in AIA 131 and (b) the plasma flows within the loops tracked in AIA 211 and 193.  The median speed is $\sim$240~km~s$^{-1}$ for the 12 AIA 131 outflows (11 down and 1 up; upflow speed $\sim$150~km~s$^{-1}$) versus a median of $\sim$1000~km~s$^{-1}$ for the 12 AIA 211 and 193~outflows.  

Figure~\ref{outflow_speeds_heights}~(c) shows initial \textit{detected} heights as a function of time as calculated using a common footpoint coordinate of (90E, 18S) as indicated in Figure~\ref{stackplot_outflows_131} (c).  (Only the 131~$\mbox{\AA}$ flows are included in this plot because their initial heights are more reliable than the plasma flows from the other filters.)  In order to point out the upward progress of these heights with time, we have overlaid a linear fit. This upward trend could be an indicator of a migrating reconnection site or the result of observational bias as the supra-arcade region becomes more visible.  The circle indicates a possible disconnection event, where an upflow and downflow appear to originate from nearly the same region simultaneously.

\clearpage

\clearpage
\newpage

\section{\label{inflows101103:discussion}DISCUSSION}

\begin{figure*}[!ht] 
\begin{center}

\includegraphics[width=.9\textwidth]{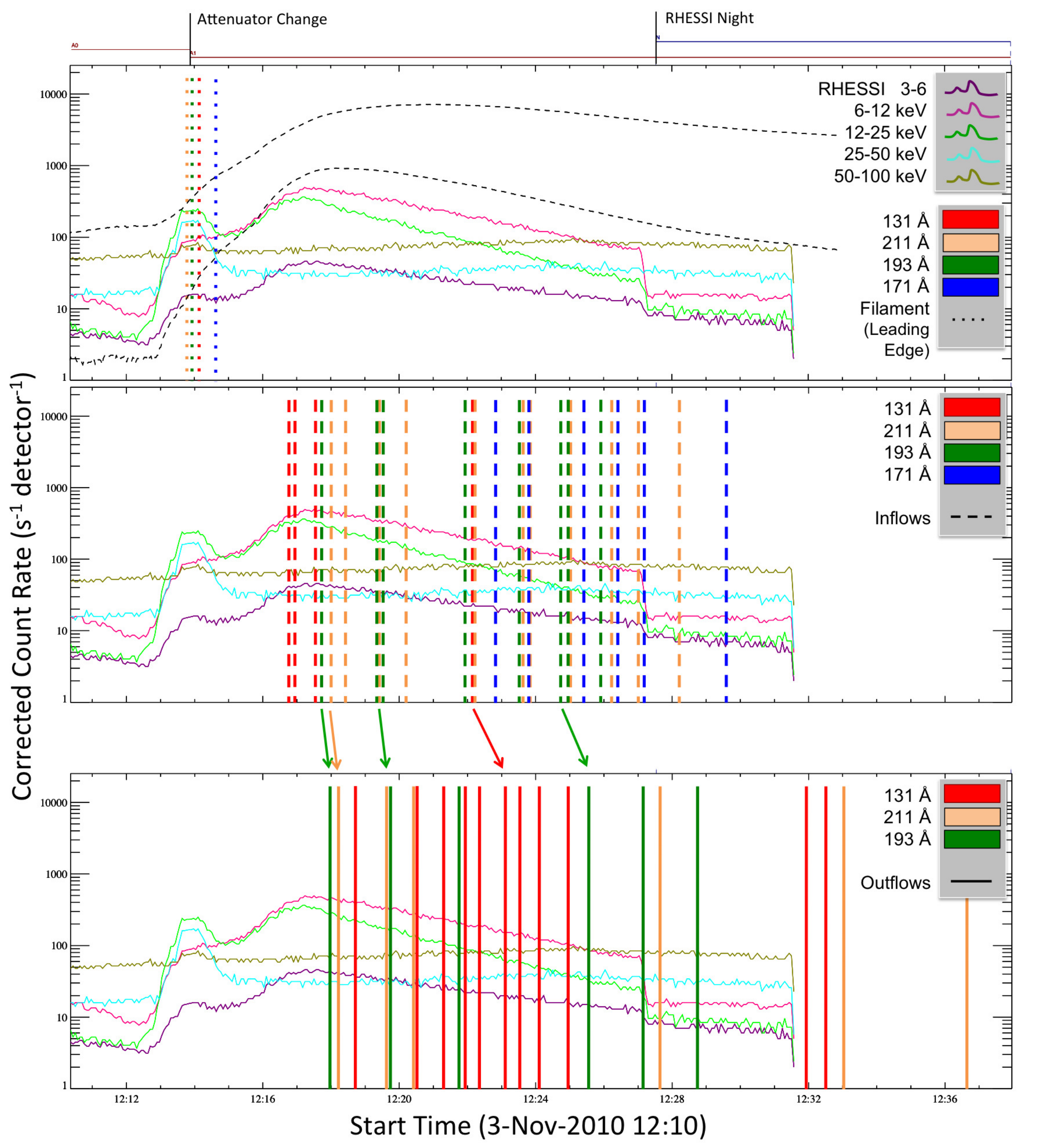}

\caption{RHESSI lightcurves (corrected count rate versus time) shown with flow times overlaid.   The shape of the GOES soft-X-ray lightcurves are shown in the top panel as dashed black lines.  The \textit{initial} detection times are plotted for the leading edge of the flux rope (top) and the outflows (bottom) while the \textit{final} detection times are shown for the inflows (middle).  The arrows indicate times of candidate corresponding inflow-outflow pairs.}
\label{lightcurve_times}
\end{center}
\end{figure*}

Figure~\ref{lightcurve_times} provides a visual representation of the flow times compared to hard-X-ray (HXR) RHESSI light curves. The initial tracked times of the leading flux rope edge (dotted) appear to occur at approximately the same time ($\sim$12:14~UT) as the first significant non-thermal peak in the RHESSI lightcurves ($\gtrsim$~25~keV).  Interestingly, this peak, which corresponds to the initial flux rope release, is not the largest.  The maximum of the HXR lightcurves occurs at $\sim$12:17~UT in the mostly thermal regime ($\la$~25~keV).  The limb obscuration of the footpoints may cause this peak sequence by reducing the non-thermal HXR flux to the detector.  \cite{glesenerEA_SPD2011} have presented findings showing a RHESSI source within the flux rope during the first peak suggesting the presence of magnetically trapped non-thermal electrons.  During the second peak, however, the RHESSI images do not indicate detectable HXR coronal emission above the compact limb source; therefore, the second peak's thermal source must be associated with increased emission near the footpoint and/or post-eruption arcade.

Note that tracking flows requires the ability to continuously observe a coherent feature within a dynamic, low signal-to-noise regime; therefore, the tracks are necessarily incomplete.  This bias leads to imprecise timings.  However, the relative timings are the more important and reliable result.  Because reconnection outflow can only occur if non-reconnected field first flows into the current layer, it is useful to compare the relative timing of the final appearance of inflows with the first appearance of outflows. Thus we have plotted the \textit{final} inflow times (dashed vertical lines) along with the \textit{initial} outflow times (solid vertical lines).

The fast, high-temperature inflows seen in AIA 131~$\mbox{\AA}$ first appear during the larger RHESSI peak and are soon followed by several outflows.  It is important to note that the determination of flow times depends strongly on observational conditions (e.g., our inability to precisely detect the final location of the inflows or the initial location of the outflows because of factors such as low inherent signal, changing morphology, and a dynamic background).  Thus it is likely that the inflows continue beyond the time we can detect them in images, and, similarly, that outflows begin before they are first seen in AIA images. Ergo, we conclude that the corresponding delay between inflow and outflow times are likely to be overestimated.  

In general we cannot discern a one-to-one correspondence between one inflow with any one outflow; therefore, it is difficult to speculate about the nature of the reconnection process itself with just the observations presented here.  However, there are five candidates for corresponding inflow-outflow pairs.   Table~\ref{table_pairs} provides information for identifying the pairs in the online movie, \textit{average} inflow speeds (the inflows experience nearly 0 acceleration over the tracked period), \textit{initial} outflow speeds (derived by fitting a 2-D polynomial to the outflow tracks), and the maximum delay time between the candidate pair constituents.  The uncertainty in the speeds for the points tracked along the inflows and the initial outflow speed measurements are approximately 20 and 50 km~s$^{-1}$, respectively.  The times of these flows are indicated by the arrows in Figure~\ref{lightcurve_times}.

\begin{table*}[!ht]
\begin{center}
\begin{tabular}{@{} c c c c c c c c @{}}
  \toprule
    {Pair} & {Wavelength}      & {Inflow} & {Outflow} & {In Speed} & {Out Speed} & {In End Time} & {Max Delay} \\
    {[\#]}   & {[$\mbox{\AA}$]} & {[\#]}       & {[\#]}         & {[km/s]}       & {[km/s]}     & {[UT]}     & {[sec]} \\
  \midrule
    1  &  131 & 8   & 12 &  260 &  310    & 12:22:09 & 60 \\
    2  &  211 & 3   &   6 &  130  & 1000  & 12:18:09 & 12 \\
    3  &  193 & 4   &   8 &  150  &  930    &12:17:45 & 12 \\
    4  &  193 & 6   & 10 &  110  &  1470 & 12:19:33 & 12 \\
    5  &  193 & 13 & 16 &  144 &  1400  & 12:24:45 & 48 \\  
  \bottomrule
\end{tabular}
\end{center}
\caption{Information concerning the candidate inflow-outflow pairs.  The inflow and outflow numbers refer to the numbered flows in the online movie.  The \textit{average} inflow speeds and the \textit{initial} outflow speeds (derived by fitting a 2-D polynomial to the outflow tracks) are provided.  The delay refers to the amount of time between the final inflow detection and the initial outflow detection.  Pairs \#2 and \#3 are flows seen in different wavelengths at the same location.  Note that the outflow of Pair \#1, a downflowing loop, is observationally distinct from the other pairs.  The other outflows appear more as traveling density enhancements along the legs of loops.}
\label{table_pairs}
\end{table*}
 
 The increase of the outflow speeds compared to the inflow speeds -- especially for the plasma moving within the legs of the loops in the 193 and 211~$\mbox{\AA}$ filters -- suggests the possibility of acceleration during the reconnection process.  The localization of the initial fast inflows, along with the short inflow-outflow delay time on order of at most a few minutes, favors fast, Petschek reconnection.  An in depth study of the magnetic fluxes and reconnection rates would be required for confirmation of such speculation.

 \clearpage

 \clearpage
\newpage
   
\section{\label{inflows101103:conclusions}CONCLUSIONS}

In this paper we have presented observations of features having characteristics (e.g., appearance, motion) consistent with the inflows predicted by standard reconnection models.  These inflows sweep into a hot, bright linear feature extending from an underlying growing arcade where their motion is promptly terminated.  We interpret the linear feature as a current sheet and provide a possible explanation for the appearance of complete loops behaving as reconnection inflows.   This work follows from previous observations reported by several authors (e.g. \citeauthor{sheeley-warren-wang_2007} \citeyear{sheeley-warren-wang_2007}; \citeauthor{narukage-shibata_2006} \citeyear{narukage-shibata_2006}; \citeauthor{yokoyamaEA_2001} \citeyear{yokoyamaEA_2001}; \citeauthor{yokoyamaEA_2001} \citeyear{yokoyamaEA_2001}), but with the addition of AIA's suite of filters and high-cadence, high-resolution capabilities.  Similar observations focusing on outflowing plasmoids along the current sheet of another limb flare are reported by \citeauthor{takasaoEA_2012} (\citeyear{takasaoEA_2012}).

An important distinction for this event's inflow observations is the addition of several resolved, loop-shaped structures retracting along the current sheet soon after the inflow motion ceases in the immediate wake of the erupting flux rope.  We have shown that the decreasing trend in inflow speed results in a similar trend in the estimated Alfv\'{e}nic Mach number.  We have also shown a correlation between the timing of the initial hot, fast inflows with a thermal peak in the RHESSI lightcurve.  Additionally, we have noted potential inflow-outflow pairs with speeds suggesting the occurrence of acceleration during the reconnection process -- as is expected from the standard models.  

A particularly interesting observation has resulted from the outflow measurements.  We have limited previous outflow tracking to flows that appeared as either SADs or SADLs because loops are directly relatable to the outflows expected from standard reconnection models.  Imaging the same region with several narrow band filters, however, allows us to observe not only the retracting loops, but possibly the traveling density enhancements within the loops.  

The loops themselves have consistently been measured to move at speeds on order of a few~$\times$~10$^{2}$~km~s$^{-1}$ \citep{savage-mckenzie_2011}, which is an order of magnitude slower than their predicted speeds based on the standard models and the assumed Alfv\'{e}n speed.  However, the apparently traveling density enhancement \textit{within} the loops, indicating either accelerated plasma, a shock, or a wave, does appear to move at speeds on order of 10${^3}$~km~s$^{-1}$.  These speeds are more consistent with doppler measurements made by \citeauthor{innes-mckenzie-wang_2003b} (\citeyear{innes-mckenzie-wang_2003b}), thereby suggesting that the outflow measurements made by those authors were of the plasma within the loop rather than the retracting loop itself.  Comparable observations from other flares, preferably with simpler magnetic topologies, need to be made to confirm or deny this assertion.

%and \citeauthor{wang-sui-qiu_2007} (\citeyear{wang-sui-qiu_2007})

But how can the retracting flux tubes be slower than the outflowing plasma?  One reasonable explanation is that the outflowing loops are slowed by field line entanglement and/or drag as they travel through the current sheet.  (The reduction in speed at the end of the flow tracks is caused by the barrier formed by the developing arcade.)  Asymmetric reconnection across the current sheet, a likely scenario for this case considering the complex morphology of the underlying field and the occurrence of low-altitude disconnection events, can also result in sub-Alfv\'{e}nic outflow speeds (\citeauthor{murphy-sovinec-cassak_2010} \citeyear{murphy-sovinec-cassak_2010}; \citeauthor{reevesEA_2010} \citeyear{reevesEA_2010}; \citeauthor{seaton_2008} \citeyear{seaton_2008}).  

While it is possible that these inflowing and outflowing motions are independent, the timing, position, and appearance of these features provide support for the case of magnetic reconnection being directly observed in the wake of an erupting flux rope (within the constraints of not having coronal magnetic field measurements).  

\acknowledgements

S. L. Savage is supported by an appointment to the NASA Postdoctoral Program at Goddard Space Flight Center administered by Oakridge Associated Universities through a contract with NASA and under the mentorship of G. Holman.   G. Holman is supported by a NASA HGI Grant and the RHESSI program.  K. K. Reeves is supported under contract SP02H1701R from Lockheed-Martin to SAO.  Support for DBS's contribution to this paper came from PRODEX grant no. C90345 managed by the European Space Agency in collaboration with the Belgian Federal Science Policy Office (BELSPO) in support of the PROBA2/SWAP mission, and from the European CommissionÕs Seventh Framework Programme (FP7/ 2007-2013) under the grant agreement no. 218816 (SOTERIA project, www.soteria-space.eu).  D.E. McKenzie is supported under contract SP02H3901R from Lockheed-Martin to MSU.  The authors would like to thank Dr. Nicholas Murphy for valuable discussions and the anonymous referee for enhancing the manuscript.

\appendix

\section{Supplemental Movie Descriptions}

\textbf{\textit{The movies will only be made available via the Astrophysical Journal after the publication date ($\sim$July 2012).\\}}

\textit{The movies provided via the online journal are intended as a guide for locating the features described in this paper.  Due to memory constraints, however, the movies are not provided at full resolution which compromises the ability to discern some of the fainter objects.}

\subsection{SECCHI - AIA Feature Matching}

A feature matching set is provided as a supplemental movie showing the 2010 November 3 event from the \textit{STEREO-B} and \textit{SDO} points of view.  All images have been reverse-scaled.  The SECCHI panels are shown using the 195~$\mbox{\AA}$ channel.  The top AIA panels are from 131~$\mbox{\AA}$ while the bottom AIA images were taken with the 193~$\mbox{\AA}$ bandpass.  The plasmas observed with the SECCHI 195~$\mbox{\AA}$ bandpass are more similar to those seen in AIA 193~$\mbox{\AA}$.  The movie spans 2.5 days as the region rotates around the eastern solar limb.  The thick dashed longitudinal line in the SECCHI images indicates the position of the solar limb as observed from Earth.

As prominent features appear throughout the event, they are highlighted in the adjacent center SECCHI and AIA images with common color annotations.  (The outer panels are the same as the inner ones but are not annotated.)  For example, the relatively small core flaring regions indicated by the red and blue dashed polygons in the SECCHI images are marked as arcade tops in the AIA images.  Limb obscuration and line-of-sight overlap cause substantial confusion within the AIA field of view.  

\subsection{Flux Rope / Inflow / Outflow Tracking}

The erupting flux rope, inflows, and outflows are tracked in the exclusively AIA 8-panel supplemental movie.  (The leading edge of the flux rope is manually tracked for each wavelength.  A description of the manual tracking procedure used for the inflows is given in Section~\ref{inflow101103:inflow_measurements}.  Finally, the apparent apexes of the descending outflowing loops as well as traveling density enhancements along loop legs, see Section~\ref{inflows101103:outflows}, are also manually tracked.)  From left to right and in general descending temperature order, the images were taken with the AIA 131, 211, 193, and 171~$\mbox{\AA}$ channels.  The top panels are the original images that have been log- and byte-scaled to enhance faint features. The bottom panels are the corresponding run-differenced versions of the top panels (i.e., preceding images have been subtracted from the current to enhance movement). The white longitudinal line indicates the eastern solar limb.  Both movies are reverse-scaled.

Features were tracked separately for each wavelength.  For the top panels, the numbers indicate the feature positions tracked within that wavelength.  The positions indicated by symbols were tracked using a different wavelength.  The corresponding symbol is located in the legend below each wavelength's panel.  (Each symbol consists of a dark and a light pattern so as to be distinguishable against the variable dark and bright background.)    The features tracked in the original images are shown only as symbols in the run-differenced images (corresponding to the symbols indicated in the top panel legends).

\end{document}